\documentclass[red,11pt,a4paper]{article}
\usepackage{amsmath}
\usepackage{amscd}
\usepackage{amssymb}
\usepackage{graphicx}
\usepackage{latexsym}
\usepackage{color}
\usepackage{floatrow}
\usepackage{float}

\title{SIMPLE MECHANICAL CUES COULD EXPLAIN ADIPOSE TISSUE MORPHOLOGY}

\author{D. Peurichard$^{1}$, F. Delebecque$^{2}$, A. Lorsignol$^{3}$, C.~Barreau$^{3}$, J. Rouquette$^{4}$, X. Descombes$^{5}$,\\ L.~Casteilla$^{3}$, P.~Degond$^{6}$}
\date{}

\begin{document}

%\begin{frontmatter}

\maketitle
\begin{abstract}
The mechanisms by which organs acquire their functional structure and realize its maintenance (or homeostasis) over time are still largely unknown. In this paper, we investigate this question on adipose tissue. Adipose tissue can represent 20 to 50\% of the body weight. Its investigation is key to overcome a large array of metabolic disorders that heavily strike populations worldwide. Adipose tissue consists of lobular clusters of adipocytes surrounded by an organized collagen fiber network. By supplying substrates needed for adipogenesis, vasculature was believed to induce the regroupment of adipocytes near capillary extremities. This paper shows that the emergence of these structures could be explained by simple mechanical interactions between the adipocytes and the collagen fibers. Our assumption is that the fiber network resists the pressure induced by the growing adipocytes and forces them to regroup into clusters. Reciprocally, cell clusters force the fibers to merge into a well-organized network. We validate this hypothesis by means of a two-dimensional Individual Based Model (IBM) of interacting adipocytes and extra-cellular-matrix fiber elements. The model produces structures that compare quantitatively well to the experimental observations. Our model seems to indicate that cell clusters could spontaneously emerge as a result of simple mechanical interactions between cells and fibers and surprisingly, vasculature is not directly needed for these structures to emerge. 
\end{abstract}

%\end{frontmatter}

\section*{Author Summary}
Because of the key role of adipose tissue in energy homeostasis and associated diseases, there is a great deal of interest in understanding the biology of this tissue. Very little is known about the key to understanding its structuration as lobules. We postulate that lobule emergence is the result of a self-organization process driven by bidirectional mechanical interactions between adipocytes and fibers. We test this hypothesis by means of a 2D individual based model of interacting adipocytes and fiber elements. Indeed, our model produces structures that compare quantitatively well to the experimental observations. This clearly shows that cell clusters of adipose tissue could spontaneously emerge as a result of simple mechanical interactions, with no direct involvement of vasculature. 

\section{Introduction}
White adipose tissue (WAT) is the main energy store of the organism. It is interconnected with all physiological functions via its endocrine functions. It plays a key role in the energy homeostasis and weight of the organism. It is a highly plastic tissue composed of differentiated adipocytes that are able to store and release fatty acids as well as to secrete numerous cytokines and hormones \cite{Ouchi_2011}. Mature adipocytes represent only 40 to 60\% of the whole cell population. The other cells form a heterogeneous population { named the stroma-vascular fraction (SVF)}. Adipocyte progenitors are present in the SVF throughout adult life \cite{Sepe_2011}. They can proliferate and/or be recruited according to physiological or pathological situations, { participate in the turnover} of adipocytes and are also believed to be supporting cells.  Because of their important role and due to the explosive worldwide development of obesity, the molecular pathways driving adipocyte differentiation are now well investigated and described \cite{Cristancho_2011}. In contrast, the global organization at the tissue scale is poorly understood. Since Wassermann's work in 1960 \cite{Wasserman_2011}, very few investigations have been performed at this scale. These seminal investigations revealed that adipose tissue is constituted of distinct lobules containing clusters of adipocytes. Moreover, observing its development, Wassermann described the emergence of mature WAT from primitive structures constituted of an unstructured fiber network containing endothelial cells and fibroblast-like cells. The latter are believed to be preadipocytes. In adult adipose tissue, lobules housing adipocytes are separated from each other by well-structured separations (or septa) composed of extracellular matrix (ECM) \cite{Napolitano_1963}. Thereafter the number of lobular units seems to remain approximately constant. In excessive development of adipose tissue occurring during obesity, increased fibrosis (formation of excess fibrous tissue) is observed and many reports associate these changes with adipocyte dysfunctions \cite{Divoux_2011, Sun_2013}. This suggests that a proper maintenance of adipose tissue architecture is critical for its normal functionality.  

Because the global architecture of adipose tissue and its organization into lobules are robust throughout adult life and { seem to be fundamental elements} of adipose tissue homeostasis, modeling the process of lobule emergence will greatly improve our understanding of adipose tissue biology and plasticity in physiological or pathological conditions. Numerous models of tissue morphogenesis can be found in the literature, describing the emergence of self-organization of cells and fibers. Due to their simplicity and flexibility, the most widely used models are Individual Based Models (IBM) (see \cite{Drasdo_2003}, \cite{Hwang_2009} and references therein). They describe the behavior of each agent (e.g. a cell or a fiber element) and its interactions with the surrounding agents over time. 
Due to the high computational cost of IBM, mean-field kinetic or continuous models, which are more efficient to describe the large scales, are often preferred. All these models include one or several of the following interactions: (i) cell/cell, (ii) cell-fiber and (iii) fiber-fiber interactions. Models of interacting cells moving in ECM-free media such as \cite{Drasdo_2005} focus on interactions of type (i). Based on the mechanisms reviewed in \cite{Friedl_2000}, a wide variety of models incorporating interactions of type (ii) have been proposed such as: (a) mechanical models \cite{Murray_1983}, (b) chemotaxis-type models (\cite{Lushnikov_2008, Ambrosi_2005} and references therein), where cell motion is driven by chemical gradients or (c) models of contact guidance \cite{Guido_1993} (see \cite{Dickinson_2000, Hillen_2006, Hillen_2010}) where the ECM gives directional information for cell motion. However, how the processes are coordinated to produce directed motion is not well understood. Fiber-fiber interactions have been explored in \cite{Alonso_2014}, where a model of a fibrous network composed of cross-linked fiber elements is proposed. Other authors treat the fibrous network as a continuum, such as a porous medium \cite{Taber_2011} or an active gel \cite{Joanny_2007} for instance. However, the literature so far provides little clues on the mechanisms underlying contact guidance or fiber self-organization. In the present paper, we demonstrate that directionally organized cell and fiber structures can emerge without appealing to contact guidance or fiber directional interactions, as a result of simple mechanical interactions between the cells and the fiber network. { To test this hypothesis qualitatively, it is sufficient in a first instance to consider a two-dimensional model, which we do for reasons of simplicity and computational efficiency.} Our model is of more microscopic nature than previously proposed mechanical models \cite{Murray_1983, Shraiman_2005} and aims at describing the emergence of the lobular structures observed in adipose tissue. 
{ Our goal is to test the scenario that,} due to the fibers resisting the pressure induced by the growing adipocytes, the latter are forced to regroup into clusters. Adipocyte clusters in turn force the fibers to merge into a well-organized network. To validate this scenario, we developed a two-dimensional IBM modelling adipocytes interacting with ECM fiber elements. The model and experiment data showed strikingly similar lobule-like structures and revealed that vasculature was not needed for lobular self-organization to emerge, although vasculature is a proven key factor of adipogenesis \cite{Corvera_2014}. 

\section{Material and methods}
\subsection{Experiments and image processing} 
\begin{figure}[H]
\centering
\includegraphics[scale=0.36]{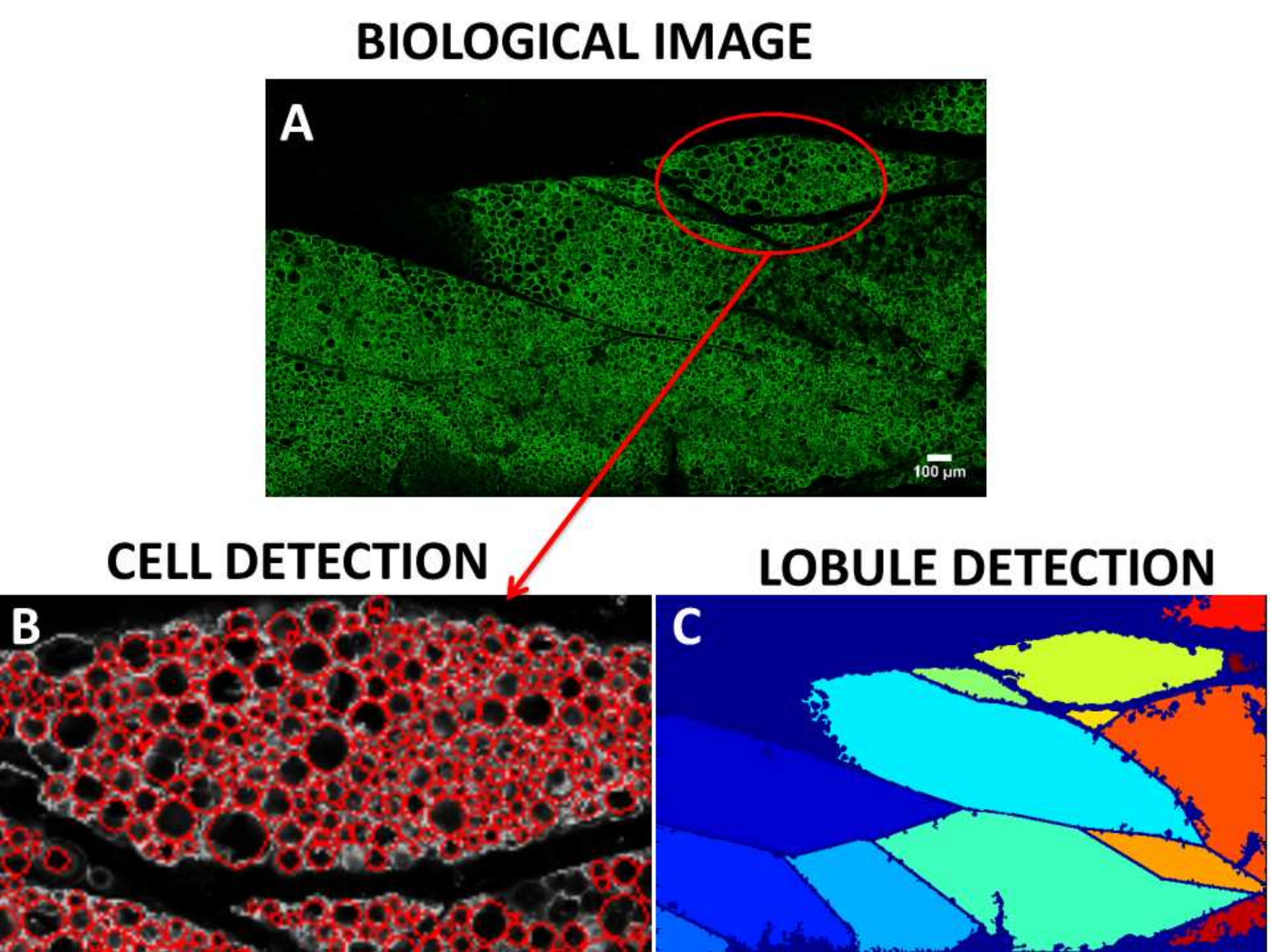} 
\caption{{\bf Adipose tissue imaging} (A) 2D Image of a part of mouse sub-cutaneous adipose tissue. Lipid droplets were immunostained for perilipin (green). ECM between adipocyte clusters appears in black. (B)~Magnification of the part enclosed by the red line on image (A), showing the result of cell detection. Cells appear as red circles. (C)~ Image (A) after lobule detection. Detected lobules have been distinguished by different colors.\label{Fig1}}
\end{figure}

Fig. \ref{Fig1} shows part of a sub-cutaneous adipose tissue from an adult mouse. The adipocytes were visualized by immuno-staining of perilipin, a protein that surrounds the unilocular lipid droplets. In Fig.~\ref{Fig1}A, the ECM fiber network appears in black as a background. This picture clearly reveals the organization of the adipose tissue in lobules. A 3D image of one isolated lobule is shown in supplementary information. We implemented classical image processing methods to extract the centers and radii of the cells (Fig. \ref{Fig1}B) and the different lobules (Fig. \ref{Fig1}C) (see \ref{App:IP} for details). The quality of the cell and lobule segmentation methods were carefully checked. 

\subsection{Description of the model}\label{Main:Model}
 We postulated that, in WAT, the agents contributing the most to mechanical balance were the ECM fibers and the adipocytes. { The ECM was discretized into unit fiber elements consisting of line segments of fixed and uniform lengths represented by their centers and their directional unit vectors. We supposed that two fiber elements that crossed each-other could form a link, thereby creating a longer fiber.} {The cells were described as 2D spheres represented by their centers and radii.} At any given time, the two sets of agents were supposed to realize the minimum of the mechanical energy of the system (described below). 
We incorporated the following biological features~: {\em (i) Pre-adipocyte differentiation:} Immature cells are much smaller than adipocytes. So, we supposed that they had negligible impact on the mechanical equilibrium and we did not incorporate them in the model. The transformation of an immature cell into an adipocyte was {modelled} as the creation (or ``insemination'') of a new adipocyte. All new adipocytes were inseminated with the same small radius. New adipocytes were inseminated at random times following a Poisson process. The location of the insemination was also random with either uniform probability in the domain {(below referred to as ``random insemination'')} or with a bias resulting in a higher insemination probability at locations where existing adipocytes were already present {(below referred to as ``biased insemination'')}. In this biased insemination case, the existing cell density in a disk of radius $R$ around the randomly chosen insemination point $X$ was computed and normalized by the maximal possible density (corresponding to adipocytes in contact with each other), resulting in a dimensionless parameter $\chi$ comprised between $0$ and $1$. Then, the insemination probability at $X$ was taken proportional to $\chi^\alpha$, with biasing parameter $\alpha >0$. { In this scenario, a pre-adipocyte sensed the adipocyte density $\chi$ up to a sensing distance $R$ and made a decision whether to differentiate into an adipocyte according to the value of the parameter  $\chi^\alpha$ (the larger $\alpha$, the larger the local adipocyte density needed to be to trigger differentiation). { By correlating the new adipocytes location to existing ones, biased insemination indirectly accounted for a vascular network bringing blood supplies in given locations of the tissue to trigger the appearance of new adipocytes (see Section 'Discussion' below)}}. {\em (ii) Adipocyte growth:} The ability to store and release energy according to the needs of the organism was modelled through the regular growth of the cells. Therefore, thanks to (i) and (ii), we incorporated both hyperplasia (cell number increase) and hypertrophy (cell size increase). As the turnover of adipocytes is small and not related to adipose tissue morphology \cite{Arner_2010}, we neglected the apoptosis of adipose cells. We assumed that the volume of each adipocyte reached a maximal value beyond which it stayed constant. {\em (iii) Adipocyte incompressibility and non-overlapping:} Adipocytes are reservoirs of fat, which is an incompressible liquid, and they cannot overlap. Therefore, we assumed that the radius of each disk was unaffected by whatever mechanical efforts were exerted onto it and that two neighboring disks could not overlap.{ { \em (iv)~Fiber resistance to adipocyte pressure:}  We viewed the ECM as a soft medium composed of fibers having locally preferred directions. Each fiber element represented the elementary mechanical action that the ECM exerted on neighbouring cells. It carried a directional information corresponding to the preferred direction of the collagen fibers. The largest mechanical action of the fiber was exerted normally to this direction. The fibers repelled the cells by means of a soft potential allowing some penetration of the cells inside the ECM. A fiber element produced a unit of potential strength. Larger mechanical actions were achieved by having several of these unit fiber elements in a close neighbourhood. Obviously, there exists a limit to ECM strength, but we assumed that our system operated below that limit and we did not implement any upper bound on the number of fiber elements per unit area. In practice, we assumed that the fiber-cell repulsion potential iso-lines were ellipses with focii at the two ends of the fiber segment and  that the potential vanished beyond a certain distance from the fiber.{ The choice of elliptic isolines corresponded to the simplest anisotropic shape in 2D}. {\em (v) Fiber growth, elongation and ability to bend:}  In addition to carrying a unit of ECM fiber strength, fiber elements also carried a unit of fiber length. However, we provided a way to create longer fibers by allowing two crossing fibers to create a cross-link. Any displacement of a cross-linked fiber pair would maintain the position of the cross-link relative to the center of each fiber. Several consecutively cross-linked fiber elements would model a long fiber having the ability to bend or even take possible tortuous geometries. As the number of fibers cross-linked to a given fiber was not limited, we could also account for fiber branching. Therefore, the cross-linking process would model fiber elongation \cite{Ciarletta_2009} and symmetrically, pairs of cross-linked fibers could also spontaneously unlink, allowing for fiber breakage describing ECM re{modelling} processes. Linking and unlinking processes followed Poisson processes with frequencies $\nu_\ell$ and $\nu_d$ respectively, and the parameter $\chi_\ell = \frac{\nu_\ell}{\nu_\ell+\nu_d}$, where $\chi_\ell \in [0,1]$ represented a measure of the fraction of linked fibers among the pairs of intersecting fibers. {\em (vi) Fiber alignment and resistance to  bending:} To model the  ability of long fibers (those made of several cross-linked fiber units) to offer a certain resistance to bending, linked fibers were subjected to a potential torque at their junction. This torque vanished when the fibers were aligned, and consequently acted as a linked-fiber alignment mechanism. Any force exerted by a cell in the vicinity of a cross-link would result in the cross-linked fibers making an obtuse angle with respect to one another until the exerted torque at the cross-linked balanced the effect of the force, thereby providing a discrete representation of the bending of a continuous beam. 
This torque was characterized by a stiffness parameter $c_1>0$ playing the role of a flexural modulus. The larger the $c_1$ the more rigid the fiber assembly was.}
 
 The mechanical energy of the system included the cell-fiber interaction potentials (iv) and the linked fiber-fiber alignment potential (vi). At each time step, a minimum of this mechanical energy subject to the nonoverlapping constraint between cells (iii) and to the linkeage constraint between linked fibers (v) was sought. At the beginning of the next time step, new adipocytes were inseminated (i), adipocyte radii were increased (ii) and pairs of fibers were linked/unlinked (v). These phenomena disrupted the mechanical equilibrium and a minimum of this new energy was again sought, and so on.  

\begin{figure}[H]
\centering
\includegraphics[scale=0.25]{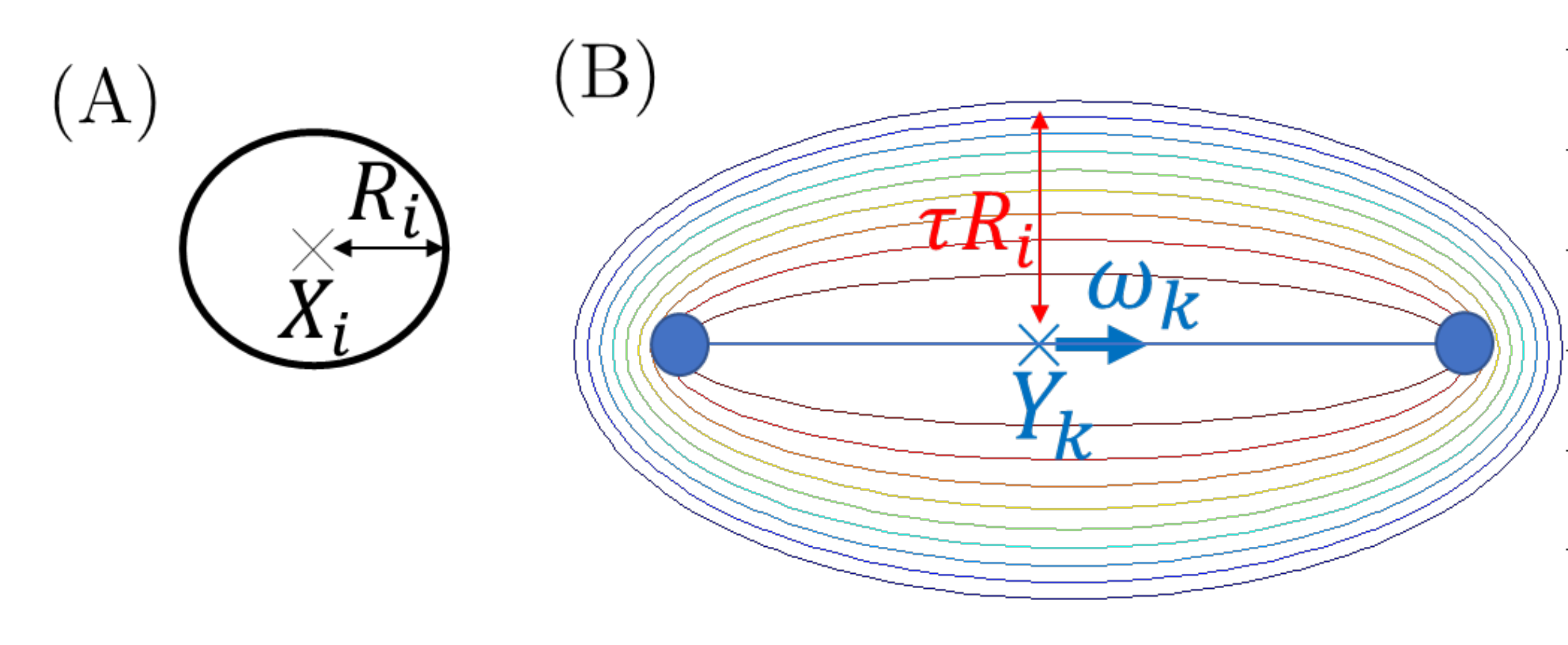}
\caption{{\bf Cell-fiber interaction potential.} A: Cell $i$ of radius $R_i$. B: Isolines of a potential generated by an horizontal fiber $k$ with $\tau=2$. Maximal repulsion distance $\tau R_i$ in the orthogonal direction of the fiber.}
   \label{potim}
\end{figure}

More specifically, the $N_A$ adipocytes were modeled as 2D growing spheres of center $X_i$ and radius $R_i$ for $i$ in $[1,N_A]$, and the $N_f$ ECM fiber elements were represented by straight lines of fixed length $L_f$, of center $Y_k$ and orientational angle $\theta_k$ for $k \in [1,N_f]$.  Given a configuration at a fixed time, we denoted by $C$ the set of cell center positions and radii: $C=\{(X_i, R_i) \; , \; i \in [1,N_A]\}$ and by $F$ the set of fiber center positions and fiber directional angles: $F = \{ (Y_k, \theta_k) \; , \; k \in [1,N_f]\}$. The global mechanical energy of the system was written: 
\begin{equation}
\displaystyle \mathcal{W}(C,F) = W_{pot} (C,F) + W_{al} (F),\label{pot}
\end{equation}
\noindent where $W_{pot}$ and $W_{al}$ were the fiber-cell repulsion potential (iv) and the linked fiber-fiber alignment potential (vi) respectively:
\begin{align}
& W_{pot} (C,F) = \sum_{1 \leq i \leq N_A} \sum_{1 \leq k \leq N_f} W_{i,k} (X_i,Y_k,\theta_k) \label{potpot}\\
& W_{al}(F) =  c_1 \sum_{ (k,m) \in [1,N_f]} p^t_{km} \sin^2(\theta_k - \theta_m).\label{potal} 
\end{align}
\noindent The time-dependent coefficients $p^t_{km}$ were set to 1 if fibers $k$ and $m$ were linked at time $t$ and to 0 otherwise (see \ref{App:timeScale}). The alignment potential $W_{al}$ of intensity $c_1$ consisted of the sum of elementary alignment potentials between fibers of a linked pair. The repulsion potential $W_{pot}$ was supposed to be the sum of two-particle potential elements, $W_{i,k}$, modeling the mechanical interaction between cell $i$ and fiber $k$. The iso-lines of these potential elements consisted of ellipses with focii located at the two ends of the fiber segment (see Fig. \ref{potim}). We supposed that  fiber $k$ could repel cell $i$ up to a distance $\tau R_i$ in its orthogonal direction, where $R_i$ is the radius of the cell and $\tau$ a model parameter. {We stress the fact that fiber-cell repulsion aimed to model repulsion of the fiber by the border of the cell, while a cell was modeled by its center. Therefore, the repulsion distance $\tau R$ of the model corresponded to a repulsion distance $(\tau - 1) R$  from the border of the cell.}
At each time step, the minimization of the global mechanical energy under the constraints (iii) and (v) was written:
\begin{equation}
\label{argmin}
(C,F) = \underset{\Phi(\tilde{C})\leq 0 , \; \Psi(\tilde{F})=0}{\textrm{argmin}} \mathcal{W}(\tilde{C},\tilde{F}),
\end{equation}
\noindent where $\Phi(\tilde{C})\leq 0$ expressed cell-cell nonoverlapping inequality constraints (iii) and   $\Psi(\tilde{F})=0$ expressed fiber-fiber linkeage equality constraints (v) (see \ref{App:Constraints}), where:
\begin{equation*}
\begin{split}
 \Phi(C) &= \big( \Phi_{ij}(X_i,X_j)\big)_{(i,j) \in [1,N_A]^2}, \\
 \Psi(F) &=  \big(\vec{\Psi}_{km}(Y_k,Y_m,\theta_k,\theta_m) \big)_{(k,m) \in \mathcal{N}_f}.
 \end{split}
 \end{equation*}
\noindent  Here, $\mathcal{N}_f$ denoted the set of linked fiber pairs: $\mathcal{N}_f = \{ (k,m) \in [1,N_f]^2 , \, k < m , \, p_{km}^t = 1\}$. We send the reader to \ref{App:Uzawa} for more details on the numerical algorithm used for solving the minimization problem. The model was implemented on a 2D square domain and boundary conditions were assumed periodic (i.e. each agent was assumed periodically repeated beyond the boundary of the square domain). The numerical simulations were initialized with a randomly distributed fiber network (according to a uniform distribution over all possible direction angles or over a sub-interval of directions angles centered about a given angle $\theta_1$ and of width $2 \theta_2$). New cells were inseminated at a constant rate until reaching a cell volume fraction of 50\%, a number consistent with the experimental observations.  We refer the reader to \ref{App:Model} for a more complete description of the model ingredients.

\subsection{Biological relevance of the model parameters}
{ The reference time was chosen to be 10 times the insemination time $t_{ins}$ (inverse of insemination rate), i.e $t_{ref} = 10 t_{ins}$, and the time step was chosen to be $\Delta t = t_{ins} = 0.1 t_{ref}$. Simulations were stopped at dimensionless time $t=100 t_{ref}$. The time to reach the maximal number of cells corresponded to $t_e=18 t_{ref}$ and that for cells to reach their maximal sizes to $t_g = 18 t_{ref}$. Simulations were run five times the time needed for all cells to reach their maximal sizes to ensure that an equilibrium was attained in the end. The fiber unlinking time $t_d = 1/\nu_d$ varied between $t_d = 10 t_{ref}$ and $t_d = 10^4 t_{ref}$, and the linking time $t_\ell = 1/\nu_\ell$ ranged from about twice to ten times $t_d$ according to the value of $\chi_\ell$. Note that these are linking/unlinking times per fiber but the actual frequencies of linking/unlinking events must be multiplied by the number of fibers $N_f$ ($N_f=800$ in our simulations, see \ref{App:timeScale} for details on the time scales). It is difficult to relate these time scales to actual biological time scales because adipocyte growth rate is highly variable according to the subject, its age, its diet, its energy consumption, etc. Nonetheless, adipocyte turnover is low \cite{Arner_2010} and so we can estimate that it takes about 100 days for a nascent adipocyte to grow to its maximum size. In our model, a nascent adipocyte needs about 20 time units to grow, so, it is consistent to assume that one time unit of our model is about five days. In our model, we can estimate that there is significant ECM remodeling when at least $10 \%$ of the fibers (i.e about 100 fibers) have been linked or unlinked. Therefore, the ECM remodeling rate can be related to $100 \, \nu_d$ and consequently, the ECM remodeling time $ t_d/100$ ranges between $0.5$ and $5 10^2$ days. In \cite{Cowin_AnnRevBiomedEng04}, it is estimated that ECM remodeling takes about 15 days which falls between these two bounds. 

{ By estimating the adipocyte radius to $30 \mu m$ from the experiments (see Fig. \ref{Fig1}), the fiber elements considered here were $60 \mu m$ large and $140 \mu m$ long. In \cite{Ushiki_2002}, it is shown that the collagen fibers are organised into bundles of collagen fibrils. The bundle width varies from $1$ to $20 \mu m$ depending on tissues and organs, and in loose connective tissues such as adipose tissue these bundles can run parallel to each other to be twined into larger bundles. As for the length of the fiber elements, we note that actual collagen fibers are able to bend around the cells. As in our model fiber elements are rigid, the bending ability of the fibers is restored by connecting fiber elements together without forcing them to be aligned (the finite value of the flexural modulus $c_1$ allows for an angle between two connected fiber elements to appear). However, if the fiber elements are too long, the curvature of the resulting fiber may be too restricted. In order to describe fibers able to bend around the cells, the fiber element length should be about a cell diameter, i.e 60 $\mu$m. In \ref{App:smallFib}, we report on simulations using the values of fiber element length of 54 $\mu$m and width 10$\mu$m consistent with these considerations. However, these values lead to computationally intensive simulations and for the purpose of building a phase diagram of the system, we instead have used fiber element length 140$\mu$m and width 60 $\mu$m that lead to tractable simulations. { If not otherwise stated, the model parameters used for the simulations are given by table \ref{tablemodparamGD}, the values of the numerical parameters are listed in table \ref{tableparamnum} and more information on the values of the parameters for the biological phenomena can be found in \ref{App:Model}.

\begin{figure}

\caption{{\bf Model parameters} \label{tablemodparamGD}}
\centering
\footnotesize
\begin{tabular}{|c|c|c|c|}
\hline
Name & Value & Dimension & Description\\
\hline
\multicolumn{4}{|c|}{\textbf{Fibers}}\\
\hline
$N_f$ & $800$ & N/A & Number of fibers\\
\cline{1-2}
$L_f$ & 3 & 136 $\mu m$& Fiber length \cite{Ushiki_2002}\\
\hline 
\multicolumn{4}{|c|}{\textbf{Cells}}\\
\hline
$N_A$ & $190$ & N/A & Maximal number of cells\\
\hline
$R$ & $[0.1, 0.66]$ & $[4,30] \mu m$ & Range of cell radii\\
\hline
\multicolumn{4}{|c|}{\textbf{Mechanical cell-fiber repulsion potential}}\\
\hline
$W_0$ & $10$ & N/A& Minimal potential force\\
\cline{1-2}
$W_1$ & $15$ & N/A & Maximal potential force\\
\cline{1-2}
$\tau R$ & 2$R$ & $60 \mu m$& Fiber-cell repulsion distance\\
\hline
\multicolumn{4}{|c|}{\textbf{Mechanical fiber-fiber alignment}}\\
\hline
$c_1$ & 1 & N/A& Fiber-fiber alignment force\\
\hline
\multicolumn{4}{|c|}{\textbf{Biological phenomena}}\\
\hline
$\frac{1}{100 \nu_d}$  & $[10^{-1}, 100]$& $[5,   500]$ days& ECM remodelling time\\
\hline
$\chi_\ell$ & 0.35& N/A& Linked fiber fraction\\
\hline
\end{tabular}
\end{figure}

}The relevance of these parameters is assessed in\ref{App:smallFib} where we observe that the results with the more biologically relevant parameter values are similar to those obtained with the larger values, thus justifying the use of the latter for an extensive exploration of the parameter space. }

\section{Results}\label{Main:Results}
\subsection{Influence of the model parameters}
 Fig.~\ref{Fig2} (I) shows a phase diagram representing the various morphologies obtained using random adipocyte insemination and varying the linking/unlinking frequency $\nu_d$ (horizontal axis of the diagram) as well as the linked fiber fraction $\chi_\ell$ (vertical axis of the phase diagram). It illustrates that the linking-unlinking dynamics strongly influenced the morphology of the final structures. Indeed, varying $\nu_d$ and $\chi_\ell$, we observed that the cell cluster morphology changed from compactly shaped (Fig.~\ref{Fig2} (I 1) or (I 4)) to elongatedly shaped (Fig.~\ref{Fig2} (I 3) or (I 6)), while the fiber cluster morphology changed from disordered (Fig.~\ref{Fig2} (I 1)) to aligned (Fig.~\ref{Fig2} (I 2)). Finally the fibers self-organized into rigid long fiber threads (Fig.~\ref{Fig2} (I 3)). For a slow linking-unlinking process (Fig.~\ref{Fig2} (I 1)), the fiber network topology was almost frozen and this generated rigidly connected fiber structures that were too stiff to self-organize (see \ref{S1_Vid} and \ref{S2_Vid}). A faster linking-unlinking process (Fig.~\ref{Fig2} (I 2)) allowed for the remodeling of the network topology and for the local alignment of the fibers thanks to the alignment torque generated at the links (see \ref{S3_Vid}).  
 \begin{figure}[H]
\centering
\includegraphics[scale=0.12]{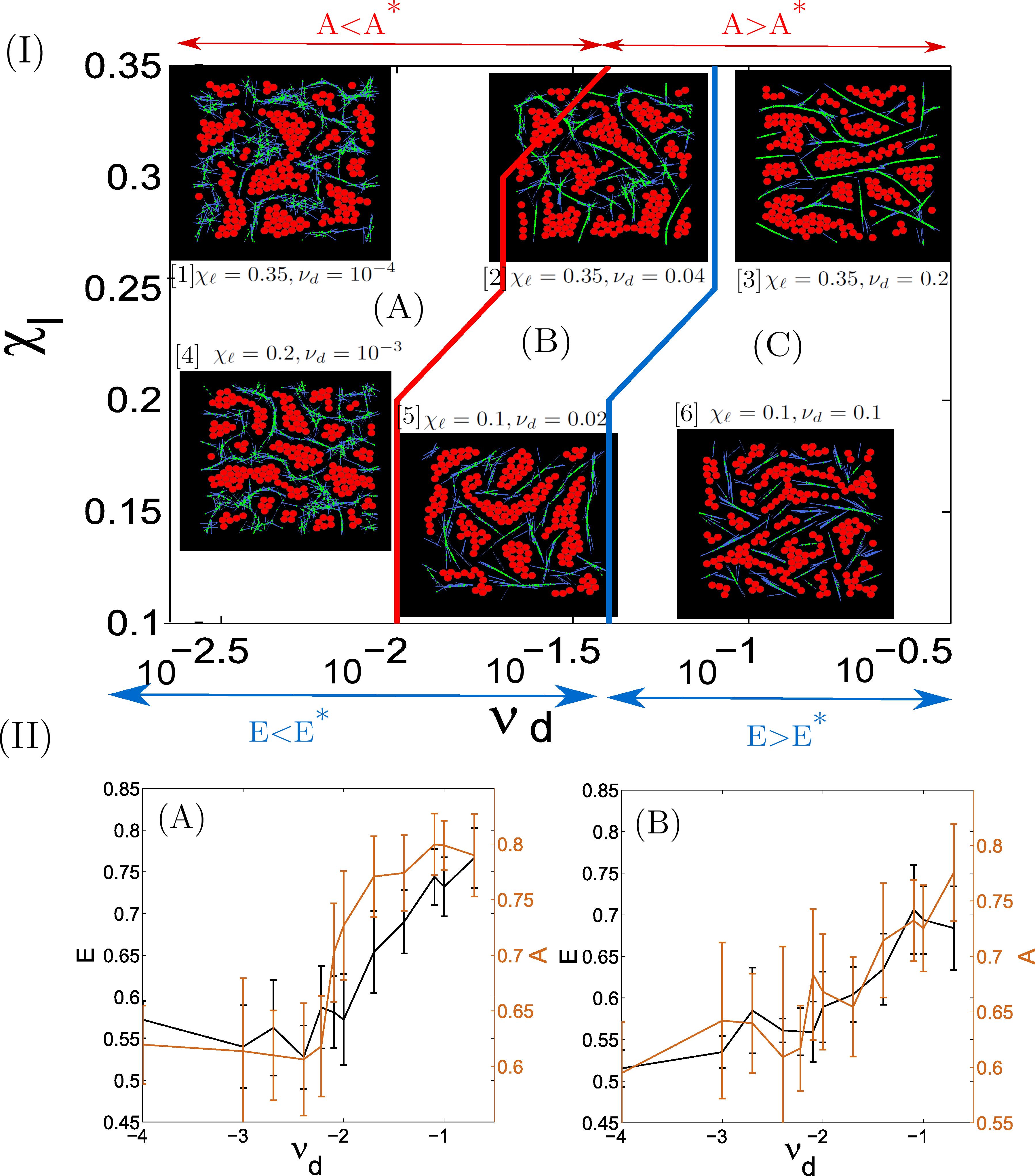}
\caption{{\bf Influence of the model parameters} (I): Phase diagram representing the
separation lines between three morphologies: (A) Lobule-like cell clusters surrounded by a disorganized (unaligned) fiber network, (B) lobule-like cell structures in an aligned fiber network, and (C) elongated cell structures in a network composed of long and rigid
fiber threads, as functions of the unlinking frequency $\nu_d$ and of the linked fiber fraction $\chi_\ell$. The passage from one morphology to another is quantified by two statistical quantifiers: the mean cell cluster elongation $E$ and the mean alignment of fiber clusters
$A$ (Phase (A): $A < A^*$ and $E < E^*$ ; Phase (B): $A > A^*$ and $E < E^*$ ; Phase (C): $A > A^*$ and $E > E^*$). The separatrix $A = A^*$ between Phases (A) and (B) is plotted in red, while the separatrix $E = E^*$ between Phases (B) and (C) is plotted in blue. For each phase, two simulations in the corresponding range of parameters $\nu_d$ and $\chi_\ell$ are shown. (II): fiber cluster mean alignment $A$ (in orange) and mean cell cluster
elongation $E$ (in black), averaged over 10 simulations and plotted as functions of the unlinking frequency $\nu_d$ for linked fiber fraction $\chi_\ell = 0.1$ (Fig. (II A)) and $\chi_\ell = 0.35$ (Fig. (II B)).The mean cell cluster elongation $E$ increases with $\nu_d$, with two plateaus for $\nu_d \leq 10^{-2}$ and $\nu_d \geq 10^{-1}$ whatever $\chi_\ell$ is. The fiber cluster alignment $A$ increases with $\nu_d$ with a plateau value $A \approx 0.6$ for $\nu_d \leq 10^{-2}$.\label{Fig2}}
\end{figure}
However, for a very fast linking-unlinking dynamics (Fig.~\ref{Fig2} (I 3)), the aligned fiber threads were reinforced by the fast creation of links, increasing the rigidity of the network (see \ref{S4_Vid}). A preferred fiber direction locally emerged and favored the growth of cell clusters in that particular direction, thereby generating elongated cell clusters. For a smaller linked fiber fraction $\chi_\ell = 0.1$ (Fig.~\ref{Fig2} (I 5) or (I 6)), the fiber network was less rigid compared to the previous case. Fiber clusters were consequently smaller, and failed to surround the cell structures, generating bigger cell clusters. 

In Fig.~\ref{Fig3}(A) and (B), we show simulation results obtained with the larger fraction of linked fibers $\chi_\ell = 0.35$ and for $\nu_d = 10^{-3}$ (A) and $\nu_d = 2 10^{-2}$ (B). Similarly, Fig.~\ref{Fig3}(C) and (D) show simulation results using biased insemination with biasing parameter $\alpha = 0.1$. Again, the larger linked fiber fraction $\chi_\ell = 0.35$ is used, and two values of $\nu_d$ are investigated: $\nu_d= 10^{-3}$ (C) and $\nu_d = 2 10^{-2}$ (D). It shows that biased insemination did not have a significant influence on the final cell and fiber structures. For a well calibrated fiber linking-unlinking dynamics, the model was able to produce lobule-like structures without the need of biased insemination (see \ref{App:EB}} for more details).

\subsection{Quantitative analysis}
In order to quantify cell and ECM fiber structures, we defined a set of statistical descriptors (SQ).  A cell cluster was defined as a group of at least $5$ adipocytes in contact with each other, and $N_C$ was the number of cell clusters per 100 adipocytes. Parameter $E$ measured the average elongation of cell clusters. It was comprised between $0$ (for disk-like cell clusters) and $1$ (for cord-like clusters) { and had the expression:
\begin{equation*}
E = \frac{\sum_{c=1}^{N_C} \textrm{Card}( \mathcal{R} \cap \mathcal{C}_c )}{\sum_{c=1}^{N_C} n_c},
\end{equation*}
\noindent where $\mathcal{R} \cap \mathcal{C}_c$ is the set of cells with less than five neighbors in cluster $c$, and $n_c$ its total number of cells (see \ref{App:SQ} for more details)}. We verified that our conclusions did not depend on the chosen minimal size (here $5$) of the cell clusters. { Due to computational constraints, we characterized cell clusters by their shape rather than by their number of cells. Indeed, comparing the number of cells with experimental data would have required to simulate a larger number of cells, which was computationally too intensive}. The SQ $\Theta$ measured the standard deviation of the shape anisotropy direction of cell clusters. For this purpose, each cell cluster was best-matched to an ellipse {(again because ellipses are the most simple anisotropic shapes in 2D)} and the cluster shape anisotropy direction was defined as the angle of the ellipse semi-major axis with a reference direction. Small values of $\Theta$ indicated a preferred shape anisotropy direction of cell-clusters. {Moreover, a fiber cluster was defined as a set of neighboring quasi-aligned fiber elements, and parameter $A$ measured the mean alignment of the fibers of the clusters. Parameter $A$ was comprised between 0 and 1: small values of $A$ indicated an isotropy in the fiber orientation angle distribution ('disorganized fiber network') while large values of $A$ indicated a global alignment of the fibers in the clusters ('organized fiber network'). { The expression of $A$ was given by:
\begin{equation*}
A = \frac{1}{N_f}\underset{1\leq c_f\leq N_{T_f}}{\sum} \lambda^{+}_{c_f} n_{c_f},
\end{equation*}
\noindent where $N_f$ is the total number of fibers, $\lambda_{c_f}^+$ is the mean alignment of fibers in cluster $c_f$ and $n_{c_f}$ its number of fibers.} We refer the reader to \ref{App:SQ} for details on the computation of the SQ.}

\subsection{Identification of different morphologies} 
{For each set of model parameters, we computed the SQ on the obtained final structures and averaged them over $10$ realizations. { In Fig.~\ref{Fig2} (II), we plotted the mean alignment of fiber clusters $A$ (in orange) and the mean cell cluster elongation $E$ (in black) as functions of the fiber unlinking frequency $\nu_d$ for two values of the linked fiber fraction $\chi_\ell = 0.1$ (Fig.~\ref{Fig2} (II A)) and $\chi_\ell = 0.35$ (Fig.~\ref{Fig2} (II B)) (see \ref{App:MainR} for a complete diagram scanning the values of $\chi_\ell \in [0.1,0.35]$ and for the SQ $\Lambda$ and $N_C$). These plots revealed an increase of the cell cluster elongation $E$ as the unlinking frequency $\nu_d$ increases, whatever value the linked fiber fraction $\chi_\ell$ took. We identified two plateaus of values of $E$: $E \approx 0.55$ for $\nu_d \leq 10^{-2}$ and $E \approx 0.75$ for $\nu_d \geq 10^{-1}$. The corresponding cell structures were compactly shaped (Fig.~\ref{Fig2} (I A) (I B), (I D) or (I E)) or elongatedly shaped (Fig.~\ref{Fig2} (I C)) respectively. Fig.~\ref{Fig2} (II) reveals that the mean alignment $A$ of fiber clusters has a plateau value for $\nu_d \leq 10^{-2}$. For a slow fiber linking-unlinking process ($\nu_d \leq 10^{-2}$), the fibers in the clusters are poorly aligned (low value of $A$), and the mean alignment of fiber clusters increases with $\nu_d$ from the value $A\approx 0.6$ to $A \approx 0.8$ as $\nu_d$ increases in $[10^{-2}, 1]$. This tends to show that the fiber linking-unlinking dynamics strongly influences the final structures. For a slow fiber linking-unlinking dynamics, due to fiber interconnections the fiber network is extremely rigid. In this case, fibers fail to align because of the high connectivity of the network which prevents any configurational change. As $\nu_d$ increases, the lifetime of each link decreases, allowing for the remodeling of the fiber network. Consequently, fiber structures are more flexible and more aligned. If the linking-unlinking process is too fast, the fiber structures easily align, forming fiber threads. These fiber threads are then reinforced by the fast creation of links, increasing the rigidity of the fiber patterns. Preferred directions locally emerge in the fiber network and favor the growth of cell clusters in these directions. The cell clusters are consequently elongated. 

We therefore identified three different morphologies -which will be referred to as 'phases'- according to the values of the parameters: (A) Lobule-like cell clusters surrounded by a disorganized (unaligned) fiber network, (B) lobule-like cell structures in an aligned fiber network, and (C) elongated cell structures in a network composed of long and rigid fiber threads. Each type of structure is obtained in a specific range of the parameters $\nu_d$ and $\chi_\ell$ of the fiber linking-unlinking dynamics. We chose the mean fiber cluster alignment $A$ and the mean cell cluster elongation $E$ to quantify the passage from one morphology to another, and we identify the threshold values $A^* = 0.68$ for $A$ and $E^* = 0.65$ for $E$. Structures of type (A) correspond to $A < A^*$ and $E < E^*$. Type (B) is described by $A > A^*$ and $E < E^*$ and finally type (C) by $A > A^*$ and $E > E^*$. Fig.~\ref{Fig2} (I) shows a phase diagram representing the various obtained morphologies in the $(E,A)$ plane. Each point in this phase diagram has been obtained by averaging the SQ over 10 simulations. The separatrix between phases (A) and (B) (i.e. the line $A = A^*$) is shown in red and the separatrix between phases (B) and (C) (i.e. the line $E = E^*$) is shown in blue. For each region, we show simulation results in the corresponding range of the parameters $\nu_d$ and $\chi_\ell$ for the sake of illustration. }
}
\begin{figure}[H]
\centering
\includegraphics[scale=0.3]{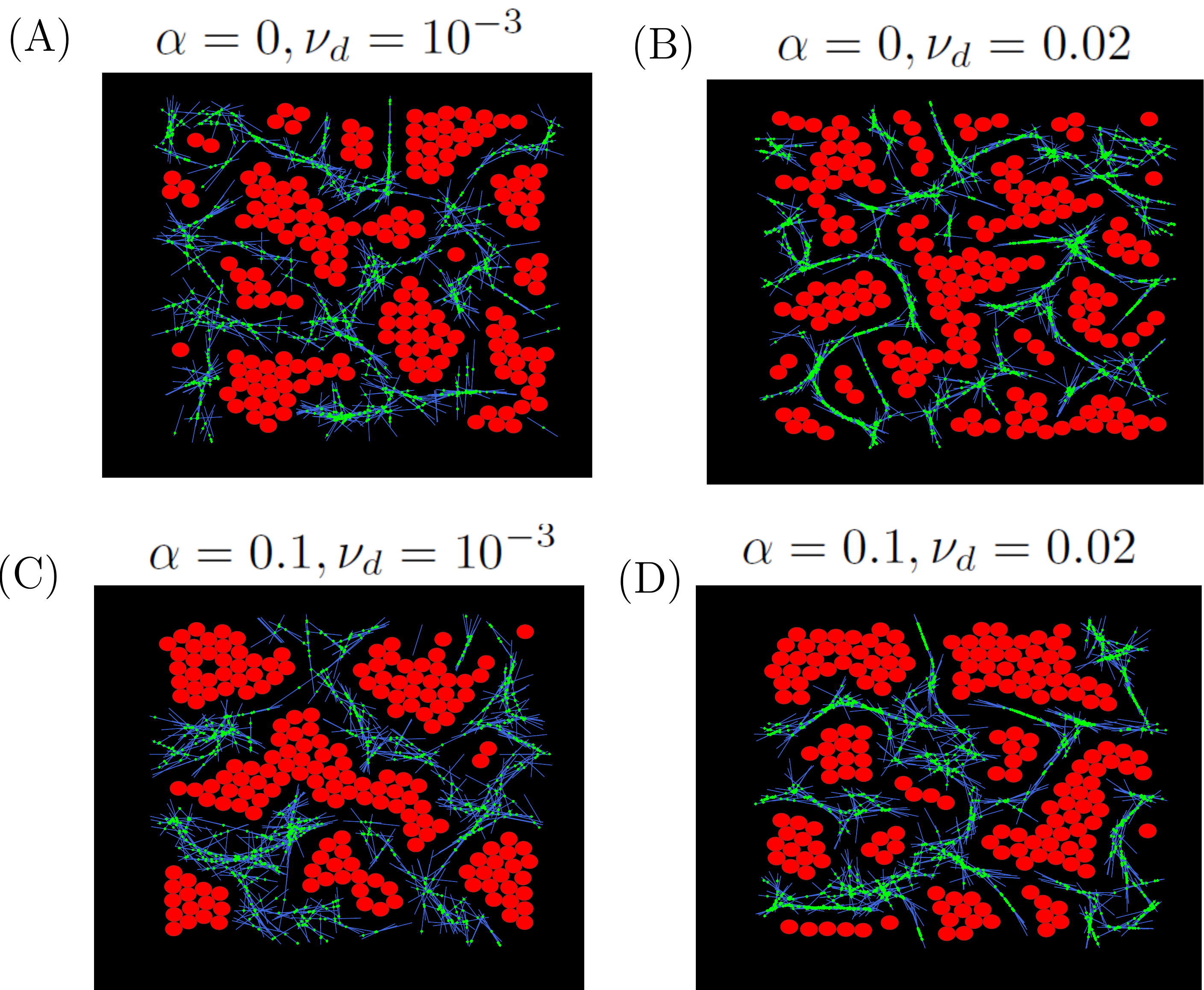}
\caption{ Model results with the larger linked fiber fraction $\chi_\ell = 0.35$ and $c_1=1$. (A) for fiber unlinking frequency $\nu_d = 10^{-3}$, (B) for $\nu_d = 2 10^{-2}$, random insemination. (C) $\nu_d = 10^{-3}$ and (D) $\nu_d = 2 10^{-2}$, biased insemination with biasing parameter $\alpha = 0.1$. Biased insemination does not significantly lead to
different morphologies compared with random insemination. \label{Fig3}}
\end{figure}

{The most biologically relevant structures, composed of well-separated lobule-like cell clusters in an organized fiber network, were characterized by a low value of the mean elongation $E$ and a moderate value of the fiber mean alignment $A$. From Fig. \ref{Fig2} I and II, we realized that the best combination of parameters was a linked fiber fraction of $\chi_\ell = 0.35$ and an unlinking frequency $\nu_d \in [10^{-2}, 10^{-1}]$. For these parameters, the model produced biologically relevant cell and fiber structures (Fig. \ref{Fig2} (I 2)), without the need for biased insemination. We refer to \ref{App:EB} for a discussion of the influence of the biasing parameter $\alpha$ and to \ref{App:c1} for the role of the alignment force $c_1$.}

In \ref{App:smallFib}, we have explored another range of parameters corresponding to those discussed at the end of Section \ref{Main:Model} above. We have shown that they give rise to similar structures and phase transitions as in the case described here. Specifically, we have used shorter fiber elements, of the order of a cell diameter. The fiber elements were made thinner in the same proportion to keep the overall shape of their interaction potential similar. Simultaneously, a decreased interaction distance between cells and fibers was used, merely reducing to a contact interaction. Finally, a fiber-fiber interaction potential, of the same strength and interaction distance as the cell-fiber interaction potential was added. To mimic fibers initially made of several fiber elements, the latter were initially placed by groups of interlinked four elements. Within this range of parameter values, we obtained very similar structures as those generated with larger fibers, larger cell-fiber interaction distance and no fiber-fiber repulsion. These results support the conclusion that the observed phenomena are generic and not tightly connected with specific parameter values. Due to computational constraints, in the simulations reported in the present section, larger values of the fiber length and width were used allowing for a smaller number of fiber elements to be simulated. This led to computationally tractable simulations which could be exploited to explore the parameter space, generate the phase diagram and calibrate the parameters for comparisons with experimental data. Finally, the convergence time to equilibrium and the influence of the final adipocyte number have been explored. Studies on the convergence time  showed that the SQ remained stable after an initial transient of less than 100 times the insemination time, which showed that the analyzed structures had correctly reached an equilibrium. Few quantitative differences were recorded when the final number of adipocytes was changed but more importantly, the important tendencies such as variations of the SQ's as functions of the model parameters remained the same whatever final adipocyte number was chosen. 

\subsection{Comparisons with experimental data} The image processing enabled the computation of the SQ $N_C$, $E$ and $\Theta$ on the biological images, thus allowing for a quantitative comparison between the model and the experimental results.  { The data were compared with simulation results at equilibrium only. Indeed, to date, the in-vivo registration and tracking of cell positions and ECM location during the development of adipose tissue is not feasible. } To compare the biological and numerical SQ, we non-dimensionalized the mean cluster number $N_C$ and the mean elongation of cell clusters $E$ by reference values referred to as $N_{ref}$ and $E_{ref}$ respectively. The numerical (respectively biological) reference values $N_{ref}$, $E_{ref}$ were defined as the mean value of $N_C$ or $E$ over all the  numerical (respectively biological) experiments. As biological images suggest that parts of adipose tissue exhibit a preferred direction, we ran simulations for different initial fiber configurations such that the initial fiber angles $\theta_{init}$ were uniformly chosen in the range $[\theta_1 \pm \theta_2]$. Parameter $\theta_1$ measured the mean initial fiber direction and $\theta_2$ was related to its standard deviation. The larger $\theta_2$, the more disordered the initial network was (with $\theta_2=\pi$ as the extreme case where the fiber initial distribution was fully isotropic). We also considered the case where the mean fiber direction could depend on the position in the form $\theta_1(x_1,x_2) = \theta_1^+$ if $x_1>0$ and $\theta_1(x_1,x_2) = \theta_1^-$ if $x_1<0$ where $(x_1,x_2)$ are the coordinates of a point in the computational domain and $x_1=0$ corresponds to the vertical middle line. This corresponded to the case where the initial fiber mean direction changed abruptly from $\theta_1^-$ to $\theta_1^+$ across the vertical middle line. 
{ We refer to \ref{App:Aniso} for a thorough analysis of the influence of the initial network anisotropy on the shape of the final structures.} Here for the purpose of comparing with actual data, we realized a database containing, for each set of model parameters $(\nu_d, \theta_1, \theta_2)$ (or in the case of position-dependent initial mean fiber direction, $(\nu_d, \theta_1^-, \theta_1^+, \theta_2)$), the SQ $N_C$, $E$ and $\Theta$ for $10$ different simulations.
The simulations were performed with random insemination and the other parameters were chosen to the value $\chi_\ell=0.35$ and $c_1=1$, according to the previous analysis. For each biological image, we first searched the database of numerical simulations to find the combination of parameters  which minimized the quadratic difference between the experimental SQ and the mean of the model SQ. Then, within the $10$ simulations generated for this set of parameters, we selected  the one minimizing the quadratic difference between the experimental and model SQ. 

\begin{figure*}
\centering
\includegraphics[scale=0.3]{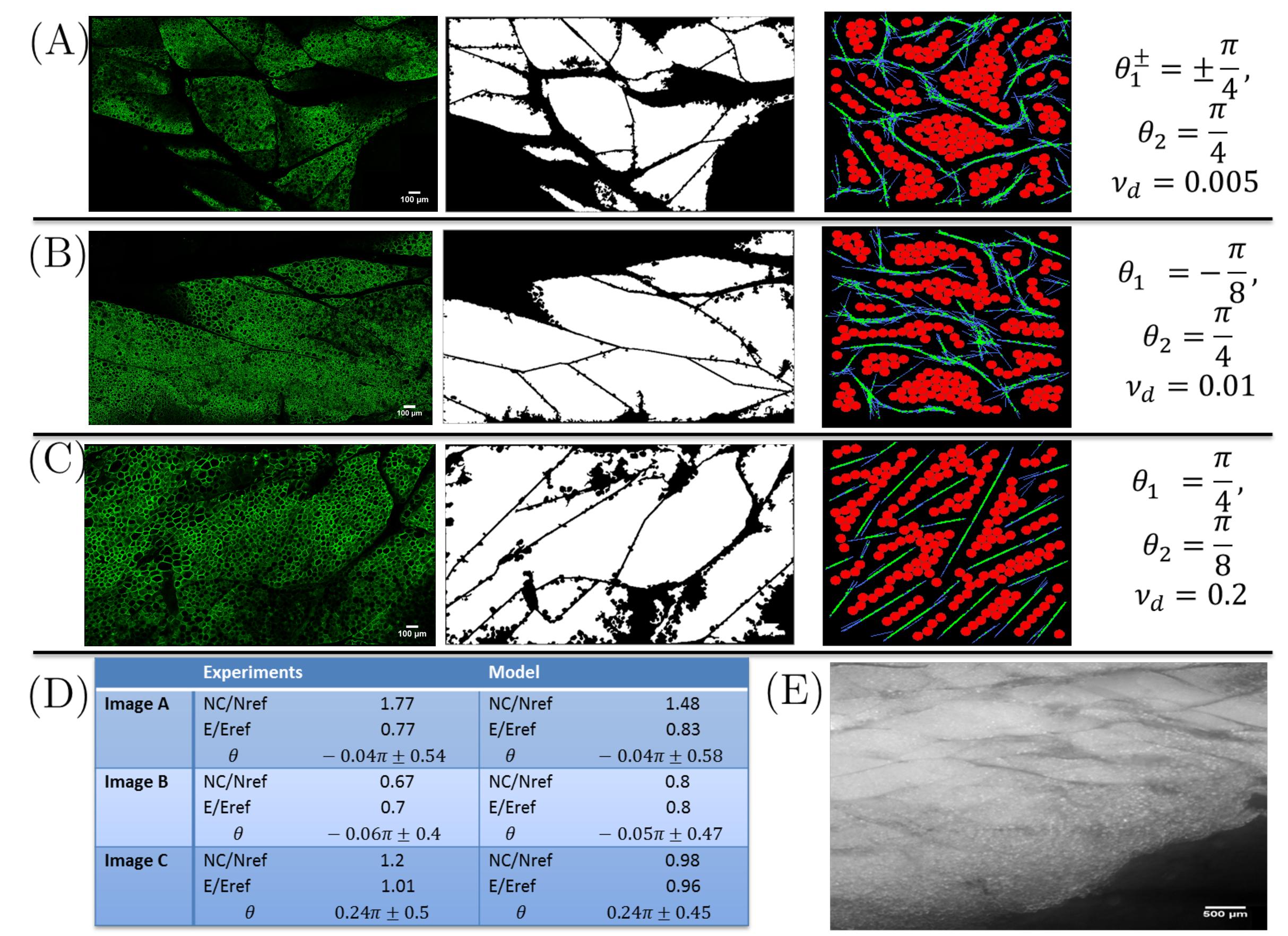}
 \caption{{\bf Comparison of the model results with experimental data}  (A) to (C)~: (I) Biological images of perilipin immunostained adipose tissue. The white scale bar at the bottom right is for $100 \mu m$. (II) Biological images after lobule detection. (III) Numerical images after simulation with parameter values offering the best correspondence with biological images. The parameter values for each simulation are indicated on the right. (D) Table showing the SQ computed on the biological and numerical images. The model reproduces the observed structures qualitatively and quantitatively well. (E): image of a large portion of adipose tissue. The white scale bar at the bottom right is for $500 \mu m$~: we notice the coexistence of similar structures as in images (A), (B) and (C) at different locations in the tissue.}
\label{Fig4}
\end{figure*}
Fig. \ref{Fig4} (A) to (C) show three biological images before~(I) and after (II) lobule detection, as well as the corresponding best simulation (III) applying the previously detailed method and the associated set of parameters $\nu_d$, $\theta_1$ (or $\theta_1^\pm$) and $\theta_2$. Table (D) provides the SQ corresponding to images (A) to (C) for both the experimental and numerical images. Finally, in Fig. \ref{Fig4} (E), a larger portion of the adipose tissue is shown (the white scale bar at the bottom right indicates $500 \, \mu m$ for (E) and $100 \, \mu m$ for (A) to (C)). 

The lobule segmentation (column (II)) revealed that the elongation of the lobular structures increased from (A) to (C). We notice that the value of the unlinking frequency $\nu_d$ corresponding to the best numerical fit increases as well, confirming the analysis made above. In these three cases, the values of the SQ given by the model are very close to the experimental ones (see Table (D)).  These results show that the model is able to reproduce the data in a fairly wide range of situations by simply modifying the model parameters. In real tissues the organizational level varies from high to low according to the position in the tissue. In Fig \ref{Fig4} (E), a large portion of adipose tissue is shown. It reveals well-organized lobular structures in its upper part and more disordered structures in its lower part. Simulation of the entire tissue using the model would be possible (although computationally intensive) by simply varying the parameters of the model to match the variation of the organizational level of the tissue. 

\section*{Discussion}
To our knowledge, this work is the first attempt to understand the emergence of the lobular structure of the adipose tissue by interfacing a mathematical model and experimental results. The originality of our work lies in the assumption that adipose tissue architecture results from a self-organization process principally driven by mechanical interactions between adipocytes and the ECM. This corresponds to a co-organization where the cell clusters and the fiber structures evolve simultaneously. Our mathematical model is able to reproduce the clustering of adipocytes into lobular units surrounded by the ECM fiber network. Simply varying a few parameters allowed us to span a large variety of morphologies. Our results suggest that adipose tissue organization could be principally driven by mechanical cues, in addition to a limited number of biologically-controlled phenomena such as fiber-fiber chemical linking. {They highlight the importance of the ECM and the mechanical forces induced by it. They are consistent with recent papers demonstrating the impact of the ECM and of its dysregulation (i.e. fibrosis) on adipose tissue function \cite{Khan_etal_MolCellBiol09, Divoux_2011, Pellegrinelli_etal_JPathol14}. 

{ Furthermore, the model can reproduce many different types of tissue morphologies by varying the parameters controlling the mechanical response of the ECM to the growth of adipocyte number and size. This gives us good confidence that although validated on mice, the model will be able to reproduce human adipose tissue. Indeed, there is increasing evidence that the differences between adipose tissue structures observed in different species or even at different locations within the same species can be related to mechanical pressure, adipocyte sizes and ECM characteristics. For instance, \cite{Sparbati_etal_EurJHistochem10} proposes to define different types of subcutaneous WAT in humans based on their observed structural and ultrastructural features. They describe lobular subtypes for fibrous white adipose tissue that can be found in regions where mechanical constraints are large. Additionally they distinguish lobular and non lobular adipose tissue on the basis of the nature and richness of the ECM. All these differences in the mechanical and structural properties of the adipocytes and the ECM can be fed into the model to produce a wide array of different morphologies. }
}
 
The structures that emerged from the mathematical model can be classified into three types: (a) middle-sized compact cell clusters surrounded by a disorganized fiber network, (b) middle-sized compact cell clusters surrounded by a well-organized network of thick fiber threads, or (c) elongated clusters surrounded by a network of thin and rigid fiber threads. Each type of structure corresponded to a range of model parameters of the fiber linking-unlinking dynamics. Structures of type (a) were obtained for a slow fiber linking-unlinking dynamics. The rigidly connected fiber structures could not self-organize, leading to a disordered fiber network. However, the system was able to produce middle sized cell clusters of lobular shape as a result of cell-fiber repulsion. This reflected the ability for a connected fiber network to exert a pressure on the cell structures and confine them into separated zones. For a faster fiber linking-unlinking dynamics, the remodeling of fiber structures was enabled, and fibers could arrange more easily into organized patterns, thanks to the torque acting on linked fibers. For a well chosen range of the unlinking frequency $\nu_d$ and of the flexural modulus $c_1$ (see discussion of $c_1$ in \ref{App:c1}), the model produced biologically relevant cell and fiber structures of type (b).  Finally, structures of type (c) were observed for fast fiber linking-unlinking dynamics. In this case, fibers easily aligned with each other and fiber stiffness was reinforced by the links and associated torque acting on linked fibers.  This imposed local directional constraints to cell cluster growth, favoring cell cluster elongation. Moreover, due to increased fiber rigidity, the fibers failed to surround the cell structures. 
{ 
The present results, which show a segregation of cells and fibers and the emergence of alignment (aka nematic order) among the fibers is consistent with experimental and theoretical work about mixtures of isotropic and nematic particles. For instance, \cite{Hosek_2004} examines a mixture of {\em F} actin and inert polymer Polyethylene glycol (PEG) and observes that actin filaments bundle together when a threshold concentration of PEG is reached. In \cite{Galanis_2010}, vibrofluidized steel rods in the presence of high concentrations of hard spheres self assemble into linear structures under similar conditions. These observations are consistent with Monte-Carlo simulations of mixtures of hard sphero-cylinders and spheres peformed in \cite{Jungblut_2007}. In these references, the coalescence and nematic ordering of the rods is analyzed in terms of the depletion force stemming from volume exclusion effects. {The phenomena observed in our simulations bear evident analogies with these observations, even though the fibers and cells do not interact via volume exclusion but through a soft repulsive potential. The difference is important in that the interaction remains  active even in the absence of noise (which was the case in our simulations) while the depletion force vanishes in the same circumstance. In particular, this explains why we observe segregation and nematic ordering in the absence of noise.} Furthermore, the aforementioned references focus on the nematic ordering of the fibers, the spherical particles being introduced only for the purpose of reducing the free volume. In the present paper, we have provided evidence that this interaction could also affect the shapes of the clusters of spherical particles (the cells) and have provided quantitative analysis of this effect. The interesting finding is that the fiber nematic order reflects itself into the elongated shape of the cell clusters, an effect which we have not seen noticed elsewhere.  }

{
By correlating the appearance of new adipocytes to a larger concentration of existing adipocytes, biased insemination equipped our model with an indirect way to investigate the influence of vasculature. Indeed, blood supplies the substrates required for adipogenesis and favors the appearance of new adipocytes at the extremities of capillaries where existing adipocytes are already present, leading to a correlation between new adipocyte locations and those of existing ones. However, fully random insemination alone reproduced the global tissue organization qualitatively well, and little improvement was observed by turning biased insemination on. This result suggests that no preferred location for  differentiation of immature cells into adipocytes is required. It looks inconsistent with Wasserman's analysis \cite{Wasserman_2011} and other works demonstrating that vasculature plays a key role in adipogenesis. Indeed, it has been demonstrated that adipose progenitors reside in the adipose vasculature and that pericytes which enwrap capillaries could represent a reservoir of adipocyte progenitors  \cite{Crisan_etal_CellStemCell08, Tang_etal_Science08}. Thus it is classically considered that the global adipose tissue architecture is primarily driven by vasculature \cite{Wasserman_2011, Corvera_2014}. Our findings open the possibility that the role of vasculature in the emergence of global adipose tissue architecture could be more complex than previously thought.  
}
 { Other phenomena were explored, such as local  fiber alignment (to model ECM reorganization by stem cells) or the suppression of isolated cells (to model isolated cell apoptosis). These additional phenomena led to a broad range of tissue structures. Although not necessarily relevant for adipose tissue, these structures could account for other organs (such as muscles, liver, etc.) or pathological adipose tissues (such as fibrotic ones). More quantitative work is needed and these questions will be developed in future works. } 
  
 Further improvements of the model could be made. { If the choice of modelling the cells as 2D spheres was motivated by our observations on the biological images (see Fig. 1), more complex cell shapes could be taken into account without profound modifications of the model. Indeed, cells could be modelled as ellipses, or as sets of several overlapping disks connected to each other. These modelling choices should have to be accompanied by appropriate expressions of the interactions/ constraints in the minimization algorithm, but the general methodology would be preserved. Moreover,  }{vasculature formation could be explicitly introduced in the model and its role as a niche for pericytes and pre-adipocytes could be explored.} Incorporating immature cells could help investigating their role in the reconstruction of the ECM \cite{Friedl_2004}. Similarly, coupling cell apoptosis with spontaneous ECM reconstruction \cite{Galle_2009} would improve the treatment of the latter. The adipocyte growth law could be made dependent on the local stress as in \cite{Chen_2001, Rodriguez_1994}. Macroscopically, introducing a disruption of the equilibrium by suppressing a part of the tissue would open up new applications of the model, such as wound healing. From a mathematical viewpoint, the development of a macroscopic model from the present IBM would allow us to perform simulations on larger domains and address the question of the organization of the whole tissue. { A first step has been made in \cite{Degond_etal_M3AS15} where a macroscopic model for interacting fibers has been derived. This macrocopic model has been further analysed and its derivation has been numerically validated in \cite{Peurichard_2016}.} From a biological viewpoint, experimental data on ECM protein deficient mice strains (such as collagen type VI) \cite{Khan_etal_MolCellBiol09} will be helpful to further validate our model.  Designing new experiments allowing for the in-vivo registration and tracking of cell positions and ECM location during the development of adipose tissue would be of great value as giving unprecedented access to the time dynamics of the morphogenesis process. 

{ { The model could also be extended to 3D without profound methodological modifications. In Fig. \ref{Fig_3Dlobule}, we show a three-dimensional image of an adipose lobule. This image reveals that while the cross-section of the lobule in a plane parallel to the skin is vaguely reminiscent of an ellipse, its shape in any plane perpendicular to the skin is much more elongated and convoluted. The results of our 2D model correctly reproduce the lobular organization in a plane parallel to the skin. A 3D extension of the model is needed to correctly account for its complex 3D organization.

\begin{figure}[h!]
\centering
\includegraphics[scale=0.9]{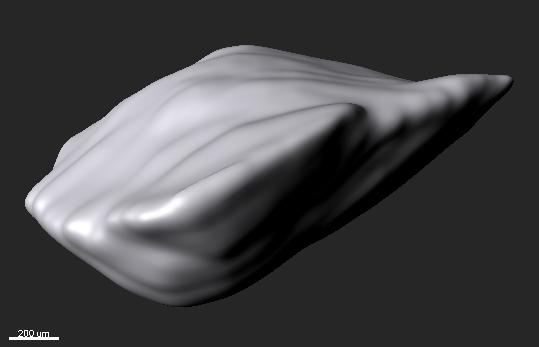}
\caption{{Three-dimensional image of an adipose lobule. The unit bar at the bottom left of the picture corresponds to $200 \, \mu$m. This image was obtained after manually delineating the lobule on  57 successive $Z$ planes (the $Z$ direction is along the elongated direction of the lobule and corresponds to the normal direction to the plane of the skin) and then reconstructing a 3D image by using the 'Imaris$^{TM}$' software}}
\label{Fig_3Dlobule}
\end{figure}

}
 To extend the model to 3D, several options are possible. Fibers could still be considered as line segments and their connection condition would allow for close but non-coplanar segments to connect with each other. Another possibility to account for the fact that fiber septa are more akin to two-dimensional surfaces would be to introduce planar ECM elements in the form of disks. Interconnecting disks would have the possibility to connect along their intersecting segments. On the other hand, cells could be modeled as spheres in 3D like in 2D, which would not require any change from the present 2D model. As mechanisms for network generation exhibit some differences in 2D and 3D, an increase of dimension in our model might affect the form of cell and fiber clusters. Testing different modelling choices for individual fibers or introducing new phenomena in the model could be necessary to reproduce the 3D structures observed in vivo, as for example to distinguish between sheet-like or tubular fiber structures according to real experiments. To this aim, more information will be needed on the biological viewpoint. When passing the model to 3D, we expect to recover that the shape of the 3D lobules is mainly controlled by the mechanical properties of the extra-cellular matrix, expectations which need to be confirmed by numerical experiments. } 

\section*{Data availability}

Data supporting this work are available on 
figshare: $https://figshare.com/projects/Simple\_mechanical$ 

$\_cues\_could\_explain\_adipose\_tissue\_morphology/16419$

\section*{Acknowledgments}
DP gratefully acknowledges the hospitality of Imperial College London, where part of this research was conducted. DP wishes to thank J. Fehrenbach (IMT, Toulouse) for enlighting discussions on image processing, and C. Guissard (StromaLab, Toulouse) for providing 3D imaging of adipose lobule.

\appendix

\section{Immunohistochemistry and confocal microscopy. }Whole mouse inguinal adipose tissues (AT) were fixed, embeded in agarose and cut into 300$\mu$m slices. Slices were permeabilized in PBS/2\% Normal Horse serum 0.2\% triton 4h at, room temperature (RT), and then incubated in anti-perilipin antibody (1/250, P1873 Sigma, 24h RT). After washing, slices were incubated in alexa488-conjugated goat anti-rabbit IgG (1/250, A11008 Molecular Probes). Imaging was performed using a Confocal Laser Scanning microscope (LSM510 NLO - Carl Zeiss, Jena, Germany) with an objective lens LCI ``Plan- Neofluar'' 25x/0,8 and excited using a 488 nm argon laser. Images were obtained by the stitching of 45 acquired images with 10\% overlapping with Fiji defined by image metadata Grid/collection stitching plugins \cite{Preibisch_2009}. Automatic image segmentation techniques were developed for (i) cell segmentation and (ii) lobule segmentation. A fully-automatic method for (i) based on a marker-controlled watershed technique \cite{Vincent_1991} was implemented (see \ref{App:IP} for details). 

\section{Mathematical model}\label{App:Model}

Here, we give details about the mathematical model presented in section \ref{Main:Model} of the main text. Let us recall the main features and introduce some notations. The two-dimensional Individual Based Model (IBM) consists of $N_A$ adipocytes, which are modeled as 2D growing spheres of center $X_i$ and radius $R_i$ for $i$ in $[1,N_A]$, and $N_f$ extra-cellular-matrix fiber elements which are represented by straight lines of fixed length $L_f$, of center $Y_k$ and orientational angle $\theta_k$ for $k \in [1,N_f]$. Adipocytes are prevented to overlap to model volume exclusion between cells. Fiber elements have the ability to link to or unlink from each other to model fiber elongation or rupture. The resistance of fibers to growing adipocytes is modeled by a repulsion potential $W_{pot}$ between cells and fibers. Additionally, fibers offer resistance to bending through an alignment potential $W_{al}$ acting between two linked fiber elements.  Cells and fibers seek to minimize their mechanical interaction energy resulting from these two potentials, subject to the non-overlapping constraint between cells and to the linkeage constraint between linked fibers. In the course of the simulation, adipocytes grow, new adipocytes appear as a result of stem-cell differentiation, new fiber links can appear and existing links can disappear. 
These phenomena disrupt the mechanical equilibrium and force the cells and fibers to move in order to restore the equilibrium.  

The outline of this appendix is the following: in \ref{App:Meca} we give the details of the mechanical interaction potentials, and \ref{App:Constraints} shows the functionals defined for the cell and fiber constraints. In \ref{App:timeScale}, we give details on the modelling of biological phenomena and detail the time scales used for the simulations. The numerical algorithm for solving the constrained problem and numerical considerations are given in \ref{App:Uzawa}-\ref{App:Comput}. 

\subsection{Mechanical interaction potential} \label{App:Meca}
 We recall that given a configuration at a fixed time, $C$ denotes the set of cell center positions and radii: $C=\{(X_i, R_i) \; , \; i \in [1,N_A]\}$ and $F$ the set of fiber center positions and fiber directional angles: $F = \{ (Y_k, \theta_k) \; , \; k \in [1,N_f]\}$. The global mechanical energy of the system reads (\ref{pot}), which we remind here for the reader's convenience: 
%\begin{equation}
$$
\displaystyle \mathcal{W}(C,F) = W_{pot} (C,F) + W_{al} (F),
%\label{pot}
%\end{equation}
$$
\noindent where $W_{pot}$ and $W_{al}$ read:
\begin{align*}
& W_{pot} (C,F) = \sum_{1 \leq i \leq N_A} \sum_{1 \leq k \leq N_f} W_{i,k} (X_i,Y_k,\theta_k) %\label{potpot}
\\
& W_{al}(F) =  c_1 \sum_{ (k,m) \in [1,N_f]} p^t_{km} \sin^2(\theta_k - \theta_m),
%\label{potal} 
\end{align*}
\noindent where $p^t_{km}$ are time-dependent coefficients that are equal to 1 if fibers $k$ and $m$ are linked and are equal to 0 otherwise. Their time-evolution is given in \ref{App:timeScale}. The repulsion potential $W_{pot}$ is supposed to be the sum of two-particle potential elements, $W_{i,k}$, modeling the mechanical interaction between cell $i$ and fiber $k$. For two given vectors $X$ and $Y$ of $\mathbb{R}^2$ and an angle $\theta \in [-\pi, \pi]$, $W_{i,k}=W_{i,k}(X,Y,\theta)$ reads:
\begin{equation}
\label{po}
 W_{i,k} = \begin{cases} \frac{\tilde{W} (\lambda^+_k)}{d_{0,i}} (d_{0,i}  - d(X,Y,\theta)) \; \text{ if $ d(X,Y,\theta) \leq d_{0,i}$}\\
0 \hspace{3,2cm} \text{   otherwise} \end{cases}
\end{equation}
\noindent where:
\begin{equation}\label{d}
d (X,Y,\theta) = |X - Y + \frac{L_f}{2} \omega(\theta)|+|X - Y - \frac{L_f}{2} \omega(\theta)|-L_f,
\end{equation}
\noindent and $\omega(\theta) = \begin{pmatrix} \cos \theta \\ \sin \theta \end{pmatrix}$ is the unit vector associated to the fiber angle $\theta$. The fiber-cell repulsion potential iso-lines are ellipses with focii located at the two ends of the fiber segment. The potential vanishes beyond distance $d_{0,i}$ to the fiber center (see Fig. \ref{potim}). Parameter $d_{0,i}$ is set such that the length of the ellipse semi-minor axis is $\tau R_i$ (see Fig. \ref{potim}), with $\tau$ a parameter set equal to $3$ in the simulations. In this case, a fiber repels cell $i$ up to a distance $\tau R_i$ in its orthogonal direction. A direct computation gives: 
\begin{equation}\label{d0}
\begin{split}
d_{0,i} = -L_f + 2 \sqrt{\Delta_{i}}, \; \Delta_{i} = (\frac{L_f}{2})^2 + (\tau R_i)^2.
\end{split}
\end{equation}
\noindent Finally, the factor $\tilde{W} (\lambda^+_k)$ in \eqref{po} measures the strength of each repulsion potential element. In order to model the fact that a fiber network is stiffer when the fibers are aligned \cite{Friedl_2000}, $\tilde{W} (\lambda^+_k)$ is assumed to be a linear increasing function of the fiber local alignment $\lambda^+_k$. The local alignment of fibers around fiber $k$ is computed in a neighborhood $B(Y_k, R_{al})$, where $R_{al}$ is the sensing distance up to which fiber $k$ senses the direction of its neighbors. Let $P_{k}$ denote the mean of the projection matrices on the direction vectors of the fibers in $B(Y_k, R_{al})$:
\begin{equation}\label{Proj}
P_k = \frac{1}{n_k}\sum_{m | Y_m \in B(Y_k, R_{al})} \omega_{m} \otimes \omega_{m},
\end{equation}
\noindent where $n_k$ denotes the number of fibers contained in $B(Y_k, R_{al})$, and $\omega_{m}$ is the directional vector of fiber $m$. The maximal eigenvalue  $\lambda^+_{k}$ of $P_k$ measures the mean alignment of the fibers in $B(Y_{k}, R_{al})$. Its corresponding normalized eigenvector gives the mean direction of the fibers in $B(Y_k, R_{al})$. A direct computation leads to:
\begin{equation*}
\lambda^+_k = \frac{1 + \sqrt{\Delta}}{2}
\end{equation*}
where 
\begin{equation*}
\begin{split}
\Delta &= 1 + \frac{4}{n_k^2}\bigg[ \bigg(\sum_{m | Y_m \in B(Y_k,R_{al})} \cos \theta_m \sin \theta_m \bigg)^2 \\
&-\sum_{m | Y_m \in B(Y_k,R_{al})} (\cos \theta_m)^2 \sum_{m | Y_m \in B(Y_k,R_{al})} (\sin \theta_m)^2\bigg] 
\end{split}
\end{equation*}
Note that $\lambda^+_{k} = 1$ when all the fibers in $B(Y_k,R_{al})$ have the same direction and $\lambda^+_{k}= 0$ when the fiber directions are fully random. The intensity of the potential element is then set to:
\begin{equation*}
\tilde{W} (\lambda^+_k) = (W_1 - W_0) \lambda^+_k + W_0,
\end{equation*}
\noindent where $W_0$ and $W_1$ are the intensities of the repulsion potential between fiber $k$ and cell $i$, when the local alignment around fiber $k$ is weak or strong respectively. 

\subsection{Constraints}\label{App:Constraints}
In order to model adipocyte incompressibility and non overlapping, we assume that the radius of each disk is unaffected whatever mechanical efforts are exerted onto it. The non overlapping constraint between cells $i$ and $j$ is written as an inequality constraint on the following function $\Phi_{ij}$: 
\begin{equation}\label{Phi}
\Phi_{ij}(X_i,X_j)=(R_i+R_j)^2 - |X_i-X_j|^2.
\end{equation}
\noindent One immediately notes that cells $i$ and $j$ do not overlap if and only if $\Phi_{ij}(X_i,X_j) \leq 0$. 

\begin{figure}
\centering
\includegraphics[scale=0.25]{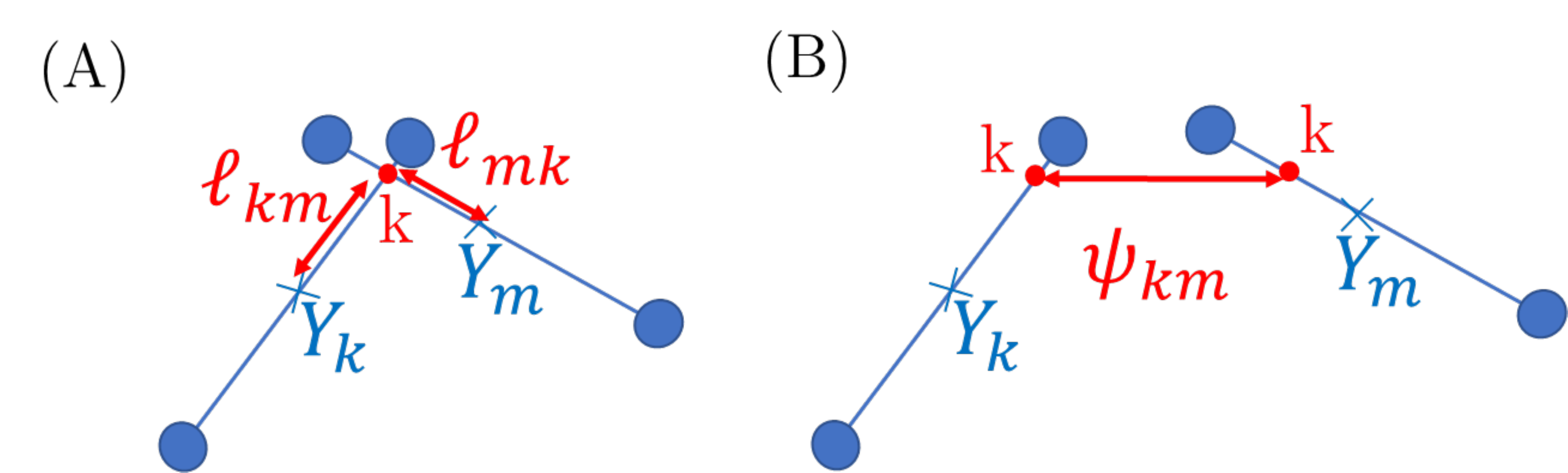}
\caption{{\bf Fiber cross-linking.} (A) Creation of a link between intersecting fibers $Y_m$ and $Y_k$. Link lengths $\ell_{mk}$ and $\ell_{km}$. (B) Constraint vector $\vec{\Psi}_{km}$ (see Eq. \eqref{P}) after fiber motion.}
  \label{fibers}
\end{figure}

To model fiber growth and elongation or conversely rupture, unlinked (resp. linked) intersecting fibers have the possibility to link (resp. unlink) at random times. As long as a pair of linked fibers remains linked, the attachment sites of the two linked fibres are kept at the same point. The maintain of the link between fibers $k$ and $m$ is modeled as equality constraints $\vec{\Psi}_{km}=0$ with:
\begin{equation}\label{P}
\vec{\Psi}_{km} (Y_k,Y_m,\theta_k,\theta_m) = Y_k + \ell_{km} \omega(\theta_k) - Y_m - \ell_{mk} \omega(\theta_m),
\end{equation}
\noindent where $\ell_{km}$ is the distance of the center of fiber $k$ to its attachment site with fiber $m$ (see Fig. \ref{fibers}) at the moment when the link is created. We use, if $\sin(\theta_m - \theta_k) \neq 0$:
\begin{equation}\label{tfk}
\ell_{km}= \frac{(x^0_m - x^0_k) \sin \theta^0_m - (y^0_m - y^0_k)\cos \theta^0_m }{\sin(\theta^0_m - \theta^0_k)}
\end{equation}
\noindent where $Y^0_k = (x^0_k,y^0_k)$ are the 2D coordinates of the center of fiber $k$ when the link is created (and similarly for fiber $m$, see Fig. \ref{fibers}). 

\subsection{Modeling of the main biological phenomena}\label{App:timeScale} 
\noindent \textit{Pre-adipocyte differentiation:} New adipocytes
of minimal radius $R_e$ are inseminated at random times following a Poisson process of frequency $\nu_{ins}$. The location of the insemination is random with either uniform probability in the domain or with a bias resulting in a higher insemination probability at locations where existing adipocytes are already present. In this last case, the probability of inseminating at a random point $X$, $\mathcal{P}(X,R)$ is a function of the cell density computed in the ball of center $X$ and of radius $R$ and normalized by the maximal possible density in this ball. This models a quorum-sensing process around point $X$ where a
pre-adipocyte senses the adipocyte density $\chi(X,R)$, up to a sensing distance $R$. This normalized density reads:
\begin{equation*}
\chi(X,R) =  \frac{\displaystyle \sum_{i | X_i \in B(X,R)} \pi R_i^2}{\pi  R_{\max}^2 N_R } ,
\end{equation*}
\noindent where $N_R$ is the maximal number of cells of radius $R_{\max}$ contained in $B(X,R)$ (for instance $N_R = 7$ for $R=2R_{\max}$). Note that $\chi \in [0,1]$. Then, the probability of inseminating at $X$, $\mathcal{P}(X,R)$, reads:
\begin{equation*}
\mathcal{P}(X,R) = \chi^{\alpha},
\end{equation*}
\noindent where $\alpha>0$ is the biasing parameter. { By denoting $t_{ins}$ the time needed to inseminate 1 cell, we define the characteristic time of the insemination process, $t_{e}$, as the mean time needed to inseminate $N_A$ cells: \begin{equation*}
t_{ins} = \frac{1}{\nu_{ins}}, \; \; t_{e} = \frac{N_A}{\nu_{ins}}.
\end{equation*}
}
\noindent \textit{Adipocyte growth}
The volumes of the cells are supposed to grow linearly with time. Given a cell $i$ at time $t$, the radius of cell $i$ at time $t+ \Delta t$ reads:
\begin{equation*}
R_i^3(t+ \Delta t) = R_i^3(t) + K_{g} (1 + \eta \rho_{g}) 
\end{equation*}
\noindent where $\eta$ is a random number chosen uniformly in $[0,1]$ and $K_{g}$, $ \rho_{g}$ are two parameters such that $\frac{K_g}{\Delta t}$ is the mean volumic cell growth per unit of time and $\frac{K_g \rho_g}{\Delta t}$ is related to the standard deviation of the volumic cell growth per unit of time.
The characteristic time of cell growth $t_{g}$ is defined as the mean time needed for a cell to reach its maximal radius $R_{\max}$ and reads:
\begin{equation*}
t_{g} = \frac{R_{\max}^3 \Delta t}{K_{g}}.
\end{equation*} 

\noindent \textit{Fiber growth or rupture:}
Fiber elongation is modeled by giving fibers the ability to attach to each other. Pairs of unlinked intersecting fibers link together at random times following a Poisson process of frequency $\nu_{\ell}$. To model fiber rupture, two linked fibers unlink following a Poisson process of frequency $\nu_{d}$.  Let $(k,m) \in [1,N_f]^2$ and define $p^t_{km}$ such that $p^t_{km}=1$ if fibers $k$ and $m$ are linked, 0 otherwise. The time evolution of $p^t_{km}$ is given by:
 \begin{equation*}
\begin{split}
 \mathbb{P}(p^{t+\Delta t}_{km}=1 | p^t_{km}=0) &= 1 - e^{-\nu_\ell \Delta t} \text{  if $\max(\ell_{km}, \ell_{mk}) \leq \frac{L_f}{2}$ }\\
 &= 0 \qquad \qquad \text{    otherwise}\\
 \mathbb{P}(p^{t+\Delta t}_{km}=0 |p^t_{km}=1) &= 1 - e^{-\nu_{d}^{km} \Delta t},
 \end{split} 
 \end{equation*}
\noindent where $\ell_{km}$ and $\ell_{mk}$ are given by Eq. \eqref{tfk}.  $\mathbb{P}(p^{t+\Delta t}_{km}=1 | p^t_{km}=0)$ describes the transition probability for a transition of $p^t_{km}$ from $0$ to $1$ during the time interval $[t,t+\Delta t]$ while $\mathbb{P}(p^{t+\Delta t}_{km}=0 |p^t_{km}=1)$ refers to the transition probability for the reverse process. We define $t_{\ell}$ and $t_d$ as the characteristic times of the linking and unlinking of fibers: respectively:
\begin{equation*}
t_\ell = \frac{1}{\nu_{\ell}},\; t_{d} = \frac{1}{\nu_{d}}.
\end{equation*}
\noindent In order to analyse the fiber linking/unlinking process, we define the ratio $\chi_\ell$:
\begin{equation*}
\chi_\ell = \frac{\nu_f}{\nu_f+\nu_d}
\end{equation*}
\noindent Note that $\chi_\ell$ is directly correlated to the fraction of linked fibers (among pairs of intersecting fibers) at equilibrium if the linking-unlinking process was acting alone and will be referred to as the 'linked fiber fraction' for short. For each unlinking frequency $\nu_d$ and each linked fibers fraction $\chi_\ell$, the linking frequency is set to $\nu_f = \frac{\chi_\ell}{1-\chi_\ell} \nu_d$.
\begin{figure}
\caption{ {\bf Model parameters} for pre-adipocyte differentiation (Poisson process) and adipocyte growth (regular process). Associated time frequencies.   \label{table3}
}
\centering
\scriptsize
\begin{tabular}{ccc}%|m{2cm}|m{2cm}|m{2cm}|}
\hline
 Phenomenon & Parameters & Frequencies\\
\hline
Cell insemination &$R=1.32$, $ R_e = 0.001$, & $\nu_{ins} = 10$ \\
&$N_A = 180$ & \\
\hline
Cell growth &$K_g = 0.0016$, $\rho_g = 0.2$, & $t_g = 18$ \\
& $R_{\max} = 0.66$ &\\
\hline
\end{tabular}
\end{figure}

\begin{figure}[h!]
\centering
\includegraphics[scale=0.18]{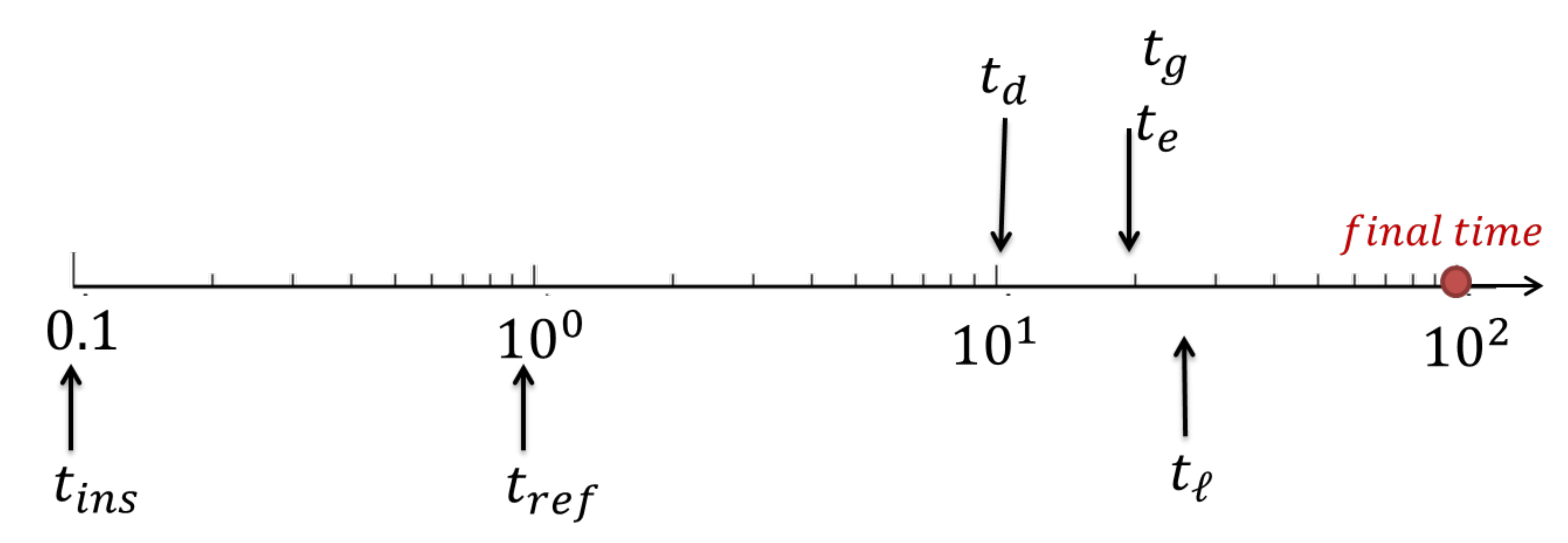}
\caption{ {\bf Time scales} for the simulations of the main text with $\chi_\ell = 0.35$ and $\nu_d = 10^{-1}$.}
  \label{timescale}
\end{figure}

{ \noindent \textit{Time scales}
The reference time is chosen to be ten times the insemination time $t_{ins}$ (inverse of insemination rate) i.e $t_{ref} = 10 t_{ins}$, and the dimensionless time step is chosen to be $\Delta t = 0.1 t_{ref} = t_{ens}$.  As explained in the main text, new cells are inseminated until reaching a cell volume fraction of 50\%. This corresponds to a total number of adipocytes $N_A = 180$. Therefore the dimensionless time after which insemination stops is $t_{e} = 18 t_{ref}$. The dimensionless mean time for a cell to reach its maximal radius is $t_{g} = \frac{R_{\max}^3 \Delta t}{K_g} \approx 18 t_{ref}$. Finally, the fiber dimensionless mean unlinking time $t_d = \frac{1}{\nu_d}$ and the ratio $\chi_\ell$ which measures the proportion of pairs of linked fibers determine the fiber linking frequency $\nu_\ell$ as follows: $\nu_\ell = \frac{\chi_\ell}{1-\chi_\ell} \nu_d$ and the dimensionless time for cross-fiber linking $t_\ell = \frac{1}{\nu_\ell} = \frac{1-\chi_\ell}{\chi_\ell} t_d$. In Fig. \ref{timescale}, we show an example of the different time scales used for the simulations of the main text, for $\chi_\ell = 0.35$ and $\nu_d = 10^{-1}$. }

\subsection{Uzawa algorithm}\label{App:Uzawa}
The constrained minimization problem \eqref{argmin} is solved with an Arrow-Hurwicz-Uzawa type algorithm \cite{Uzawa_1958}. To perform the minimization, we introduce a Lagrangian: 
\begin{equation}\label{Lsin}
\begin{split}
\mathcal{L}(C,F,\lambda,\vec{\mu}) &= \mathcal{W}(C,F) +\Phi_{\lambda}(C) 
+ \Psi_{\vec{\mu}}(F),\\
 \Phi_{\lambda}(C) &= \sum_{1\leq i,j\leq N_A} \lambda_{ij}  \Phi_{ij}(C)\\
\Psi_{\vec{\mu}}(F) &= \sum_{_{ (k,m) \in \mathcal{N}_f}} \vec{\mu}_{km} . \vec{\Psi}_{km}(F),
\end{split}
\end{equation}
\noindent where $\lambda = \big(\lambda_{ij}\big)_{(i,j) \in [1,N_A]^2}$, $\lambda_{ij}>0$ and $\vec{\mu} = \big(\vec{\mu}_{km}\big)_{ (k,m) \in \mathcal{N}_f}$,  $\vec{\mu}_{km} \in \mathbb{R}^2 $ are the sets of Lagrange multipliers of the constraints.

 Given a configuration  $(C(t_n), F(t_n))$ at time $t_{n} = n \Delta t$  the configuration $(C(t_{n+1}), F(t_{n+1}))$ at time $t_{n+1}=(n+1) \Delta t$ is defined as the limit as $p \rightarrow \infty$ of the iterative sequence $(C^p, F^p)$ where $(C^p, F^p, \lambda^p, \vec{\mu}^p)$ is defined for all $(i,j)\in [1,N_{a}]^2$ and all $(k,m)\in \mathcal{N}_f$ by: 
\begin{align}
\label{Uzawa}
&   X_i^{p+1} = X_i^p - \alpha^i_a \nabla_{X_i} \mathcal{L}(C^p,F^p,\lambda^p,\vec{\mu}^p)\nonumber\\
&   Y_k^{p+1} = Y_k^p - \alpha^k_f \nabla_{Y_k} \mathcal{L} (C^p,F^p,\lambda^p,\vec{\mu}^p)\nonumber\\
&   \theta_{k}^{p+1} = \theta_{k}^p - \alpha^k_\theta  \partial_{\theta_k} \mathcal{L}(\mathcal{C}^p,\mathcal{F}^p,\lambda^p,\vec{\mu}^p)\\
&   \lambda_{ij}^{p+1} = \max(0,\lambda^p_{ij}+\lambda_1 \Phi_{ij}(C^{p+1})), \nonumber\\
&   \vec{\mu}_{km}^{p+1} = \vec{\mu}^p_{km}+ \mu_2 \vec{\Psi}_{km}(F^{p+1})\nonumber.
\end{align}
\noindent with initial condition $(C^0, F^0, \lambda^0, \vec{\mu}^0) = (C(t_n), F(t_n),\lambda^0, \vec{\mu}^0)$ and $\lambda_{ij}^0=0$, $\vec{\mu}_{km}^0=0$, for all $(i,j) \in [1,N_A]^2$ and all $(k,m) \in \mathcal{N}_f$. 
The parameters $\lambda_1$ and $\mu_2$ control the actualization of the constraints and their choice is detailed in the next section. The minimization steps $\alpha^i_a$, $\alpha^k_f$ and $\alpha^k_\theta$ control the elementary motion of cell $i$ and fiber $k$ and the elementary rotation of fiber $k$ respectively. Their computation is detailed in the next section. The convergence test of the algorithm reads:
\begin{equation}
|\frac{\mathcal{L}^{p+1} - \mathcal{L}^p}{\mathcal{L}^p}| \leq  \epsilon_r,
\end{equation}
\noindent for a chosen $\epsilon_r>0$. Here, $\mathcal{L}^p$ is the value of the Lagrangian at iteration $p$. Because of the non convexity of the minimization problem, the uniqueness of the solution to \eqref{argmin} is not ensured and a configuration at each time step corresponds to a local minimum of the minimization problem. 

\subsection{Choice of the numerical parameters}
 In this section, the numerical parameters $\alpha^i_a, \alpha^k_f$ and $\alpha^k_\theta$ of \eqref{Uzawa} are chosen such that the amplitude of the change of each variable remains controlled. Given three bounds $\delta_a$, $\delta_f$ and $\delta_\theta$, the goal is to ensure $|X_i^{p+1}-X_i^p| \leq \delta_a$,  $|Y_k^{p+1}-Y_k^p| \leq \delta_f$ and  $|\theta_k^{p+1}-\theta_k^p| \leq \delta_\theta$, for each cell $i$ and fiber $k$. Using \eqref{Lsin} and \eqref{Uzawa}, the following expressions hold:
\begin{align*}
&|X_i^{p+1}-X_i^p| = \alpha^i_a|\nabla_{X_i} \mathcal{W}  + \nabla_{X_i} \Phi_{\lambda} |\\
&|Y_k^{p+1}-Y_k^p| = \alpha^k_f |\nabla_{Y_k} \mathcal{W}  +\nabla_{Y_k} \Psi_{\vec{\mu}}|\\
&|\theta_k^{p+1}-\theta_k^p| = \alpha^k_\theta |\partial_{\theta_k} \mathcal{W} + \partial_{\theta_k} \Psi_{\vec{\mu}}|.
\end{align*}
 \noindent The parameters $\alpha^i_a$, $\alpha^k_f$ and $\alpha^k_\theta$ are consequently set such that:
\begin{equation}\label{alpha}
\begin{split}
\alpha^i_a &= \frac{\delta_a}{2} \min (\frac{1}{|\nabla_{X_i} \mathcal{W}|},\frac{1}{|\nabla_{X_i} \Phi_\lambda|} )\\
\alpha^k_f &= \frac{\delta_f}{2} \min(\frac{1}{|\nabla_{Y_k} \mathcal{W}|},\frac{1}{|\nabla_{Y_k} \Psi_\mu|})\\
\alpha^k_\theta &= \frac{\delta_{\theta}}{2} \min(\frac{1}{|\partial_{\theta_k} \mathcal{W}|},\frac{1}{|\partial_{\theta_k} \Psi_\mu|}.
\end{split}
\end{equation}
\noindent The gradients of the potential $\mathcal{W}$ have the following upper bounds for all $i \in [1,N_{a}]$ and $k\in [1,N_{f}]$ (see \eqref{pot}-\eqref{d}):
\begin{align*}
 &|\nabla_{X_i} \mathcal{W}| \leq \sum_{1 \leq k \leq N_f} \frac{W_1}{d_{0,i}} \sim \frac{W_1 n^f_i}{d_0} \\
 &|\nabla_{Y_k} \mathcal{W}| \leq \sum_{1 \leq i \leq N_A} \frac{W_1}{d_{0,i}} \sim \frac{W_1 n^a_k}{d_0} \\
 &|\partial_{\theta_k} \mathcal{W}| \leq \sum_{1 \leq i \leq N_A} \frac{L_f W_1}{2d_{0,i}} + 2 \hspace{-.5cm} \sum_{m \; | \; (k,m) \in \mathcal{N}_f} \hspace{-.5cm} c_1 \sim \frac{L_f W_1 n^a_k}{d_0} + 2 \ell^f_k c_1 ,
\end{align*}
\noindent where $n^f_i$, $n^a_k$ and $\ell^f_k$ denote the number of fibers interacting with cell $i$, the number of cells interacting with fiber $k$ and the number of fibers linked to fiber $k$ respectively. Here, $d_0$ is the value of $d_{0,i}$ evaluated with $R_i=R_{\max}$. The following upper bounds for the gradients of the constraint functions (see Eqs. \eqref{Phi}-\eqref{tfk}) are chosen:
\begin{align*}
&|\nabla_{X_i} \Phi_\lambda |\leq 4 R_i \sum_{j \neq i} |\lambda_{ij}|,\\
&|\nabla_{Y_k} \Psi_\mu |\leq \sum_{m \; | \; (k,m) \in \mathcal{N}_f} |\vec{\mu}_{km}|,\\
&|\partial_{\theta_k} \Psi_\mu | \leq \frac{L_f}{2} \sum_{m \; | \; (k,m) \in \mathcal{N}_f} |\vec{\mu}_{km}|.
\end{align*}
\noindent These three gradient bounds are estimated at each iteration of the minimization algorithm and are included into Eqs. \eqref{alpha} to compute the values of the numerical steps.  We now turn towards the determination of $\lambda_1$ and $\mu_2$ of Eqs. \eqref{Uzawa}. Dimensionnally, the following estimations can be set from the expression of the Lagrangian Eq. \eqref{Lsin} (for all pairs $(i,j)$ and $(k,m)$):
\begin{equation*}
\lambda_{ij} = O(\frac{W_0}{R^2_{\max}}), \;  |\vec{\mu}_{km}| = O(\frac{W_0}{L_f}).
\end{equation*}
\noindent From the actualization of the constraints given by iterations of Eqs. \eqref{Uzawa}, we set:
\begin{equation*}
\lambda_{ij}  \sim \lambda_1 R^2_{\max}, \; \vec{\mu}_{km} \sim \mu_2 L_f
\end{equation*}
\noindent Then, parameters $\lambda_1$ and $\mu_2$ are set to:
\begin{equation*}
\lambda_1 \approx \frac{W_0}{R_{\max}^4}, \; \mu_2 \approx \frac{W_0}{L_f^2}
\end{equation*}
\noindent The values of the parameters $\delta_a$, $\delta_f$ and $\delta_\theta$ are taken of the order of $10^{-3}$ and the convergence test tolerance to $\epsilon_r = 10^{-5}$ (the complete set of the numerical parameters can be found in Table \ref{tableparamnum}).

\begin{figure}
\caption{{\bf Numerical parameters} \label{tableparamnum}}
\centering
\begin{tabular}{|c|c|c|}
\hline
Name & Value & Phenomenon\\
\hline
\multicolumn{3}{|c|}{\textbf{Domain}}\\
\hline
$N_s$ & 100& Total number of numerical boxes \\
\cline{1-2}
$L_s$ & $3.6$ & Length side of a numerical box \\
\cline{1-2}
$x_{\max} - x_{\min}$ & 36 & Length side of the square domain\\
\hline
$\Delta t$ & $0.1$ & Time step\\
\hline
\multicolumn{3}{|c|}{\textbf{Parameters of the minimization algorithm}}\\
\hline
$\delta_a$ & $3 10^{-3}$ & Adipocyte maximal displacement\\
\cline{1-2}
$\lambda_1$ & $30$ & Lagrange multiplier actualization\\
\hline 
$\delta_f$ & $3 10^{-3}$ & Fiber maximal displacement\\
\cline{1-2}
$\delta_{\theta}$ & $3 10^{-3}$ & Fiber maximal angular change\\
\cline{1-2}
$\vec{\mu}_{2}$ & $3$ & Lagrange multiplier actualization\\
\hline
$i_{\max}$ & 2500 & Iteration number\\
\hline
$\epsilon_r$ & $10^{-5}$ & Convergence test tolerance\\
\hline
\end{tabular}
\end{figure}

\subsection{Decreasing the computational time}\label{App:Comput}
The simulations are performed on a 2D-domain $\Omega = [x_{\min},x_{\max}] \times [y_{\min},y_{\max}]$. In order to reduce the computational time, the domain is divided into sub-squares whose side length $L_s$ is a measure of the maximal distance of the agent interactions. The goal is to compute each interaction potential element with the agents located in neighboring sub-squares of the domain only. The procedure is classical and details are omitted. Periodic boundary conditions are set by creating ghost numerical boxes of length $L_s$ at each boundary of the domain.

\section{Statistical quantifiers}\label{App:SQ}
\begin{figure}
\centering
\includegraphics[scale=0.25]{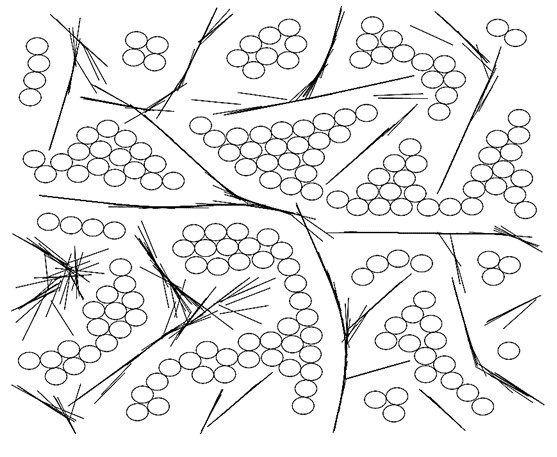}
\includegraphics[scale=0.25]{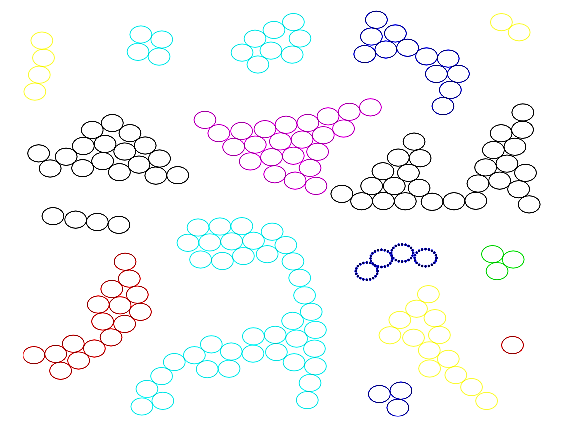}
\caption{{\bf Example of cell cluster detection.} Left: original numerical image. Cells are represented as 2D black spheres, fibers as black segments. Right: After cell cluster detection. Cells which belong to the same cluster are indicated with the same color.}
  \label{Cellclust}
\end{figure}

This section is devoted to the computation of statistical quantifiers used to describe cell and fiber structures in both numerical simulations and biological images. A cell cluster is defined as a set of cells almost in contact. Let $\sim_a$ be the reflexive and symmetric relation: 
\begin{equation*}
j \sim_a i \; \Leftrightarrow j \in \mathcal{N}_i,
\end{equation*}
\noindent where $\mathcal{N}_i$ is the set of cell $i$ neighbors:
\begin{equation}\label{ni}
\mathcal{N}_i = \{j \in [1,N_{a}] \; , j \neq i ,  \; | \; |X_i-X_j| \leq (R_i + R_j + \epsilon_a)^2\},
\end{equation}
\noindent where $\epsilon_a$ is the maximal allowed distance up to which two cells not in contact are defined as neighbors and is set to $50\% \max(R_i,R_j)$. The equivalence relation $\sim_A$ then reads:
\begin{equation*}
\begin{split}
j \sim_A i \Leftrightarrow & \exists n \in \mathbb{N}^*, \exists (a_1..a_n)\\
& \; \text{such that } j \sim_a a_1 \sim_a ... \sim_a a_n \sim_a i.
\end{split}
\end{equation*}
\noindent  Cells $i$ and $j$ belong to the same cluster if and only if $i \sim_A j$. Fig.~\ref{Cellclust} shows an example of cell cluster separation.
The statistical quantifier $N_C$ is defined as the total number of cell clusters which have more than 5 adipocytes per 100 adipocytes.

The statistical quantifier $E$ measures the mean elongation of the cell clusters, and is defined as the number of cells at the boundary of the clusters  normalized by the total number of cells in the clusters. As the parameter $E$ is irrelevant for clusters with less than 5 cells, its computation is restricted for clusters $c$ such that $n_c >5$ and reads:
\begin{equation*}
E = \frac{\sum_{c=1}^{N_C} \textrm{Card}( \mathcal{R} \cap \mathcal{C}_c )}{\sum_{c=1}^{N_C} n_c}.
\end{equation*}
\noindent Here, $\mathcal{C}_c$ is the set of indices of the cells belonging to cluster $c$, $n_c$ is the number of cells in cluster $c$ and $\mathcal{R}$ is the set of indices of all cells with less than 5 neighbors: 
\begin{equation*}
\mathcal{R} = \{ i \in [1,N_A],\  \textrm{Card}(\mathcal{N}_i) \leq 5 \},
\end{equation*}
\noindent where $\mathcal{N}_i$ is defined by Eq. \eqref{ni}.

Finally, in order to determine if the cell clusters have anisotropic shape with a preferred direction, we define the SQ $\Theta^c$ as the angle of cluster $c$ shape anisotropy direction.  For this purpose, let $X^c$ be the center-of-mass of cluster $c$, i.e:
\begin{equation*}
X^c = \frac{1}{n_c} \sum_{i\in \mathcal{C}_c} X_i.
\end{equation*}
\noindent Then, we define $P^c$ as the mean of the projection matrices on the vectors $X_i - X^c$, for all $i$ in cluster $c$:
\begin{equation*}
P^c = \frac{1}{n_c} \sum_{i\in \mathcal{C}_c} (X_i - X^c) \otimes (X_i - X^c).
\end{equation*}
\noindent The maximal eigenvalue $\lambda^+_c$ of $P^c$ gives a measure of the shape anisotropy of  cell cluster $c$ and its associated eigenvector $u^c = (u_1^c,u_2^c)$ gives the shape anisotropy direction. Then, $\Theta^c$ is defined as:
\begin{equation*}
\Theta^c = \arctan(\frac{u_2^c}{u_1^c}).
\end{equation*}
\noindent Note that $\Theta^c \in [-\frac{\pi}{2}, \frac{\pi}{2}]$. The SQ
$\Theta$ is then defined as the circular standard deviation of all the angles $\Theta^c$ for all clusters:
\begin{equation*}
\Theta = \sqrt{-2 \ln(\bar{R})},
\end{equation*} 
\noindent where $\bar{R}$ reads:
\begin{equation*}
\bar{R} = \frac{\displaystyle \sqrt{\bigg(\sum_{c=1}^{N_C} \cos \Theta^c \bigg)^2+\bigg(\sum_{c=1}^{N_C} \sin \Theta^c \bigg)^2} }{N_C}.
\end{equation*}
\noindent Finally, the mean $\bar{\Theta}$ of $\Theta^c$ over all the cell clusters reads:
\begin{equation*}
\bar{\Theta} = \frac{1}{2} \arg \big(\sum_{c=1}^{N_C} e^{2i \Theta^c}\big),
\end{equation*}
\noindent which ensures that $\bar{\Theta} \in [-\frac{\pi}{2},\frac{\pi}{2}]$. Note that large $\Theta$ corresponds to fully isotropic cell cluster organization, while small $\Theta$ indicates that cell clusters have a preferential direction.

\begin{figure}
\centering
\includegraphics[scale=0.3]{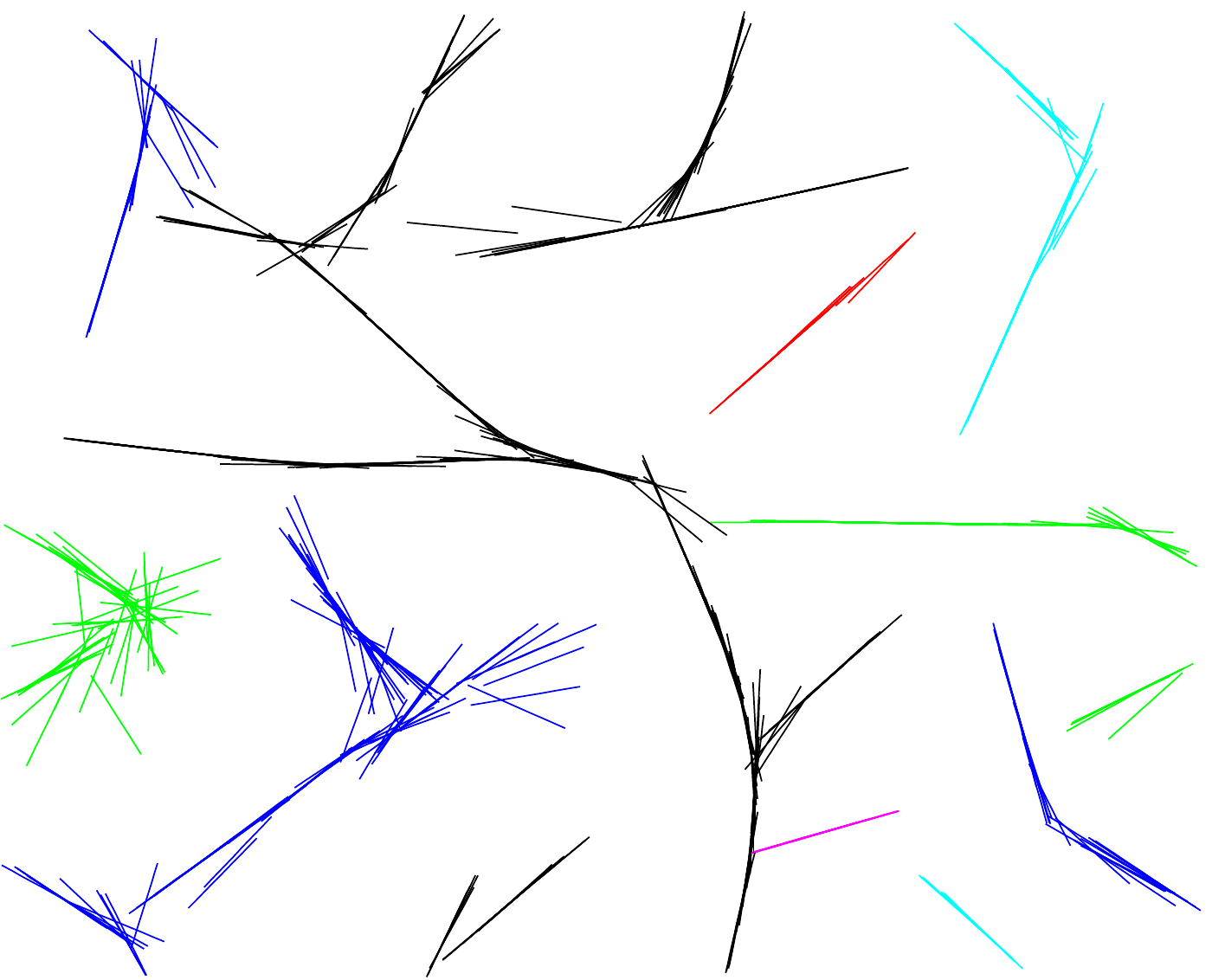}
\caption{{\bf Example of fiber cluster detection}  corresponding to simulation of Fig. \ref{Cellclust}. Fibers which belong to the same cluster are indicated with the same color.}
 \label{Fiberclust}
 \end{figure}

In order to describe the fiber structures, we define a fiber cluster as a set of neighboring quasi-aligned fiber elements and $\Lambda$ as an estimate of the average curvilinear length of such fiber clusters. Finally, $A$ measures the mean alignment of the fibers of a cluster. For this purpose, let us define $\mathcal{M}_k$ as the set of neighbors of fiber $k$, quasi-aligned with fiber $k$. Then:
\begin{multline*}
\mathcal{M}_k= \{m \in [1,N_{f}] \; , m \neq k ,\\  
 \min(d_{f}(Y_k, Y_m),d_{f}(Y_m, Y_k)) \leq 0 \\
  \text{ and } |\sin(\theta_k - \theta_m)|  < \sin(\frac{\pi}{4}) \},
\end{multline*}
\noindent where $d_f(Y_k, Y_m)$ reads:
\begin{equation*}
d_f(Y_k,Y_m) = d(Y_k,Y^-_m) + d(Y_k,Y^+_m) - 2 \sqrt{(\frac{L_f}{2})^2 + (\tau_f L_f)^2}.
\end{equation*}
\noindent Here, $ Y^\pm_m = Y_m \pm \frac{L_f}{2}\omega_m$ and $d(X,Y)$ is the distance of point $X$ to point $Y$. Note that $d_f(Y_k,Y_m) \leq 0$ (resp. $d_f(Y_m,Y_k) \leq 0$) if the center of fiber $k$ (resp. $m$) is contained in the ellipse of focii $Y^\pm_m$ (resp. $Y^\pm_k$) with semi minor axis of length $\tau_f L_f$ and semi major axis of length $\sqrt{(\frac{L_f}{2})^2 + (\tau_f L_f)^2}$. We chose $\tau_f = \frac{1}{3}$, which means that a fiber detects a neighboring fiber up to a distance $\frac{L_f}{3}$ in its orthogonal direction. This allows us to define fiber clusters as sets of quasi-aligned neighboring fibers. Let us define the reflexive and symmetric relation $\sim_f$ by: 
\begin{equation*}
k \sim_f m \; \Leftrightarrow m \in \mathcal{M}_k.
\end{equation*}
\noindent Define the equivalence relation $\sim_F$ such that:
\begin{equation*}
\begin{split}
k \sim_F m \Leftrightarrow & \exists n \in \mathbb{N}^*, \exists (a_1..a_n)\\
& \; \text{such that } k \sim_f a_1 \sim_f ... \sim_f a_n \sim_f m.
\end{split}
\end{equation*}
\noindent Then, we say that fibers $k$ and $m$ belong to the same cluster if and only if $m\sim_F k$. Fig.~\ref{Fiberclust} shows the results of fiber cluster detection from the numerical image displayed on Fig.~\ref{Cellclust}.
The mean elongation of fiber clusters is estimated by $\Lambda$. Given a fiber cluster $c_f$ and a division of the simulation domain into squares of side length $L_f$, the length of $c_f$ is estimated by $L_f \Lambda^{c_f}_F$, where $\Lambda^{c_f}_F$ is the number of squares which contain the centers of at least one fiber of $c_f$. Then, the dimensionless mean fiber cluster elongation $\Lambda$ is defined as the mean of $L_s \Lambda^{c_f}_F$ over all the fiber clusters, normalized by the maximal cell diameter:
\begin{equation*}
\Lambda = \frac{L_f}{2R_{\max} N_{T_f}}\underset{1\leq c_f\leq N_{T_f}}{\sum} \Lambda^{c_f}_F,
\end{equation*}
\noindent where $N_{T_f}$ is the total number of fiber clusters. The longer the fiber clusters, the larger the $\Lambda$. 

Finally, we define the SQ $A$ to quantify the mean alignment of the fibers of a cluster. Given a fiber cluster $c_f$, the mean alignment of its fibers is defined as the maximal eigenvalue $\lambda^+_{c_f}$ of the mean projection matrix defined by Eq. \eqref{Proj}, where the set $B(V_k,R_{al})$  is replaced by the set of all fibers of cluster $c_f$. Then, $A$ is defined as the mean of the fiber cluster alignment, weighted by the number of fibers in the cluster:
\begin{equation*}
A = \frac{1}{N_f}\underset{1\leq c_f\leq N_{T_f}}{\sum} \lambda^{+}_{c_f} n_{c_f},
\end{equation*}
\noindent where $n_{c_f}$ is the number of fibers in cluster $c_f$.

\section{Image processing}\label{App:IP}
This section is devoted to the algorithms and results of the image processing. The goal is to develop segmentation techniques for (a) cell detection and (b) cell cluster detection, in order to compute the SQs on biological images and compare them to those of numerical simulations. It is noteworthy that adipocytes only are visualized in biological images at hand, therefore SQs $E$, $N_C$ and $\Theta$ only are accessible from biological images. 

\textit{(a) Detection/separation of cells} First, a fully-automatic method for cell detection based on marker-controlled watershed segmentation has been realized.We use Marker-controlled watershed segmentation, according to the following procedure: 
\begin{itemize}
\item[(i)] The biological image is first filtered by a local median filter which associates to each pixel its mean value in its local neighborhood
\item [(ii)] The segmentation function is defined as the gradient of the transformed image. The gradient is high at the borders of the objects and low inside the objects. 
\item[(iii)] Compute foreground markers. These are connected blobs of pixels within each of the objects. The morphological techniques 'opening-by-reconstruction' and 'closing-by-reconstruction' are used to clean up the image. These operations create flat maxima inside each object that can be located using the intrinsic Matlab function imregionalmax.
\item[(iv)] Compute background markers. These are pixels that are not part of any object. We perform a simple thresholding of the intensity image: each pixel whose intensity is lower than the mean intensity of the image is set to 0.
\item[(v)] Modify the segmentation function so that it only has minima at the foreground and background marker locations.
\item[(vi)] Compute the watershed transform of the modified segmentation function.
\end{itemize}
 Object boundaries are located where $W = 0$, where $W$ is the watershed transform of the marked image gradient. This method enables the separation of multiple objects. Each object is characterized by a center (center of mass of the detected region) and a radius $R$ (radius of a circle which has the same area $a$ as the object): $R$ is thus computed as $\sqrt{a/\pi}$.

\textit{(b)Detection/separation of lobule-like clusters}. For cell cluster detection, a semi-automatic method has been developed. Each sub-images (squares occupying $0.1 \%$ of the image area) is filtered by median filtering with the intrinsic function medfilt2 of Matlab. Each output pixel contains the median value in the 3-by-3 neighborhood around the corresponding   pixel in the input image. A thresholding of the intensity image fixed at $40 \%$ of the mean intensity of the subimage is then applied. The connected objects with 8-connectivity are computed using the Matlab intrinsic function bwlabel. Finally, objects containing less than 2000 pixels (noise objects) are suppressed with the intrinsic Matlab function bwareaopen. If neighboring cell clusters are still visually connected at a point, a line is plotted by hand to separate the two clusters. The process of cell cluster detection is semi-automatic in this sense, but this procedure is sufficient for the purpose of this work, given the low number of biological images to be treated.

\section{Results and their analysis}

\subsection{Simulations of the main text}\label{App:MainR}
\begin{figure}
\centering
\includegraphics[scale=0.48]{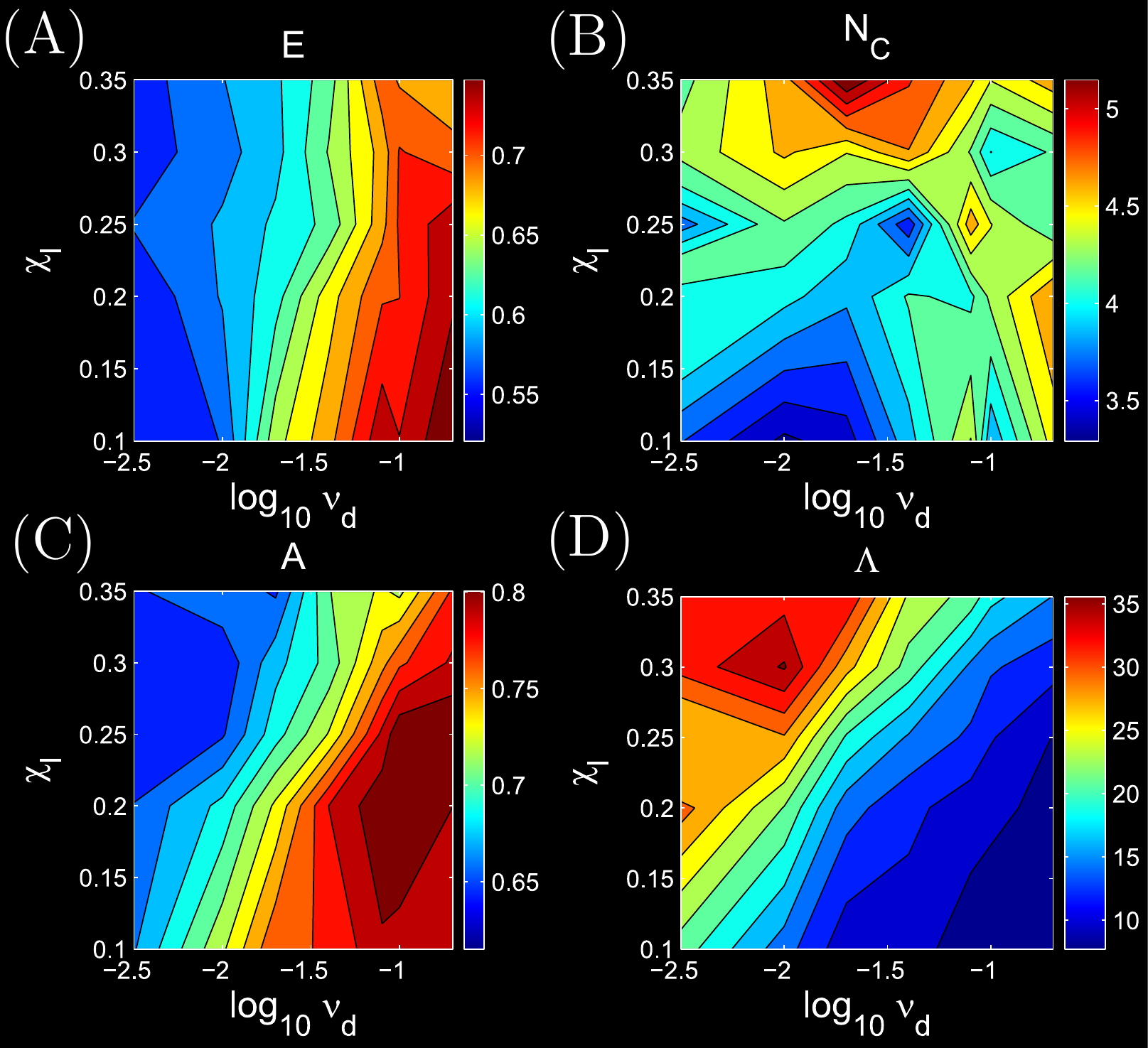}
\caption{{\bf Influence of the fraction of linked fiber pairs $\chi_\ell$ and fiber unlinking frequency $\nu_d$.}  Diagrams for (A) the cell cluster elongation $E$ ,(B) the number of cell clusters $N_C$, (C) the fiber cluster mean alignment $A$ and (D) the fiber cluster mean elongation $\Lambda$ of the simulations of the main text (see Fig. \ref{Fig2} of section \ref{Main:Results} of the main text), with random insemination, flexural modulus $c_1=1$, as functions of the fiber unlinking frequency $\nu_d$ (horizontal axis) and the linked fiber fraction $\chi_\ell \in [0.1, 0.35]$ (vertical axis). For large $\chi_\ell$, the mean fiber cluster elongation $\Lambda$ is a non monotonous function of the unlinking frequency $\nu_d$, that first increases and then decreases with $\nu_d \in [10^{-3},0.1]$.  For a value of the linked fiber fraction $\chi_\ell = 0.1$, no significant change in the mean cluster number $N_C$ arise as the unlinking frequency $\nu_d$ increases. For $\chi_\ell = 0.35$, the mean cluster number $N_C$ increases with increasing unlinking frequency $\nu_d$ in the range $[10^{-3},0.05]$ and then stay constant for larger values of $\nu_d$.
\label{mainSIMUS}}
\end{figure}

Here, supplementary results on the simulations of the main text are given. In Fig. \ref{mainSIMUS} (A and C), we show a complete diagram for the values of the Statistical Quantifiers $E$ and $A$ as functions of the fiber unlinking frequency $\nu_d$ ($x$-axis) and of the linked fiber fraction $\chi_\ell \in [0.1, 0.35]$ ($y$-axis), with random insemination and flexural modulus $c_1=1$ (see Fig. \ref{Fig2} of section \ref{Main:Results} of the main text). In Fig. \ref{mainSIMUS} (B and D), we also show the values of  $\Lambda$ and $N_C$ corresponding to the simulations of the main text. The values of the SQ are averaged over 10 simulations and we refer the reader to section \ref{Main:Results} of the main text for the analysis of the { quantifiers $E$ and $A$ and focus here on the SQ $N_C$ and $\Lambda$}. 
{ Figs. \ref{mainSIMUS} (B and D) first reveal that the fiber mean elongation $\Lambda$ and the number of cell clusters $N_C$ have plateau values for $\nu_d \leq 10^{-3}$. For a slow fiber linking-unlinking process ($\nu_d \leq 10^{-3}$), the fibers in the clusters are poorly aligned (low value of $A$, see section \ref{Main:Results} of the main text) and the mean fiber cluster elongation $\Lambda$ is fairly large, meaning that the fibers keep their initial entanglement. For large $\chi_\ell=0.35$ and as $\nu_d$ increases, the mean fiber cluster elongation $\Lambda$ increases until reaching a maximal value for $\nu_d \approx 0.005$. Then, $\Lambda$ loses $50\%$ of its value when $\nu_d$ increases in the range $[0.005,0.1]$. Indeed, as explained in Main text, if the linking-unlinking process is too fast, the fiber structures easily align but they are also more sensitive to the compression by the cells and consequently, the fiber elements regroup into shorter fiber clusters, thereby decreasing the fiber length. For a value of the linked fiber fraction $\chi_\ell = 0.1$, no significant change in the mean cluster number $N_C$ arise as the unlinking frequency $\nu_d$ increases (Fig. \ref{mainSIMUS} (B)). By contrast, for a value of the linked fiber fraction $\chi_\ell = 0.35$, the mean cluster number $N_C$ increases with increasing unlinking frequency $\nu_d$ in the range $[10^{-3},0.05]$ and then stays constant for larger values of $\nu_d$ (Fig. \ref{mainSIMUS} (B)). These results show the ability of a connected fiber network to encompass well separated cell clusters for a well chosen fiber linking/unlinking dynamics.
}

\subsection{Influence of the flexural modulus $c_1$}\label{App:c1}
\begin{figure}
\centering
\includegraphics[scale=0.23]{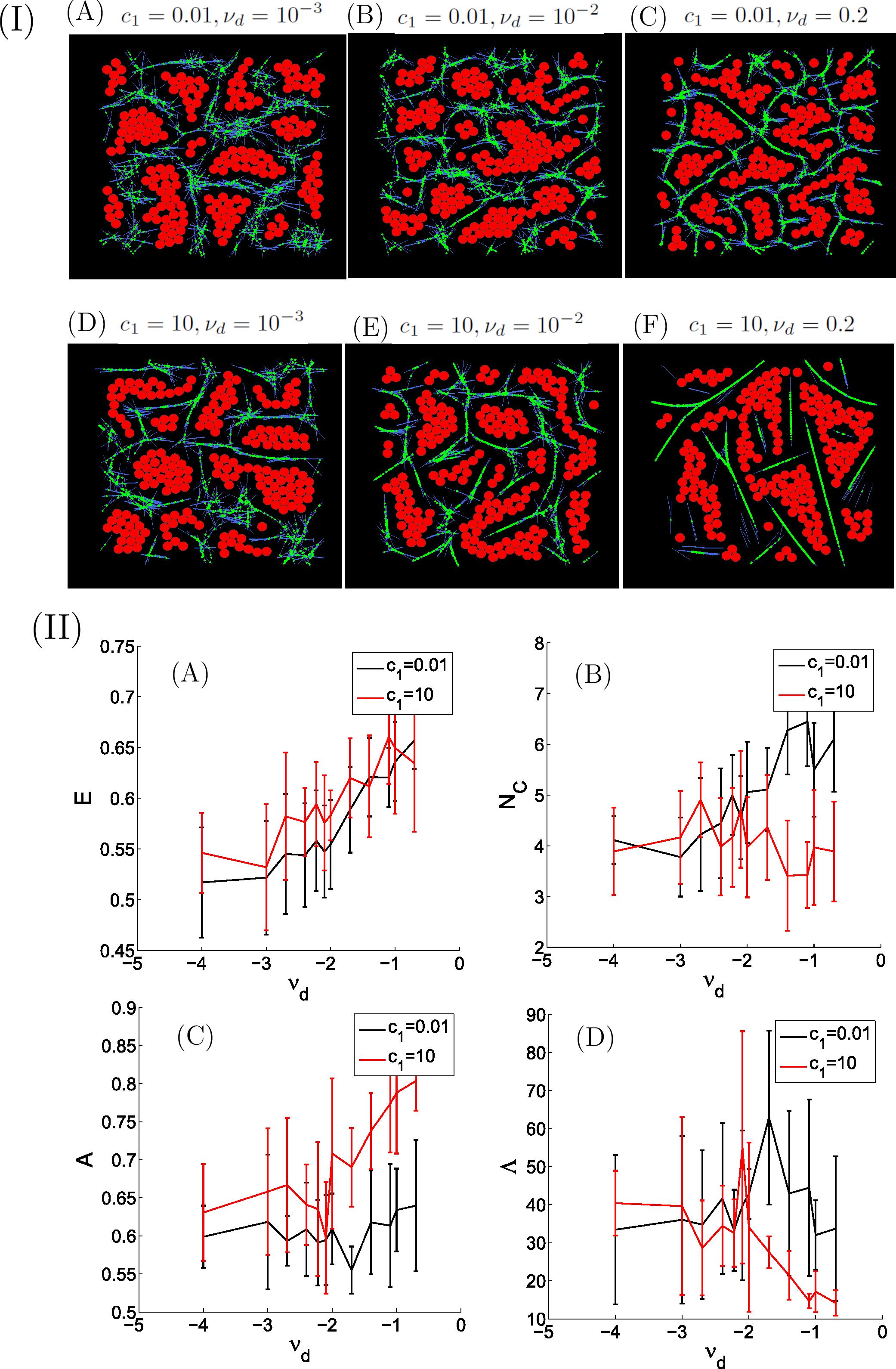}
\caption{{\bf Influence of the flexural modulus $c_1$.} (I) Simulations with random insemination and two different flexural moduli $c_1 = 0.01$ (first row) and $c_1 = 10$ (second row), for linked fiber fraction $\chi_\ell = 0.35$ and three different unlinking frequencies $\nu_d = 10^{-3}$ (Figs. (I A) and (I D)), $\nu_d=10^{-2}$ (Figs. (I B) and (I E)) and $\nu_d = 0.2$ (Figs. (I C) and (I F)). For a small flexural modulus $c_1 = 0.01$, fiber structures change slowly from disorganized clusters (Figs. (I A) and (I B)) to more aligned fiber patterns (Fig. (I C)) as $\nu_d$ increases. A larger value of $c_1 = 10$ leads to more aligned fiber clusters. The morphology changes from well-organized fiber clusters surrounding separated cell clusters  (Figs. (I D) and (I E)) to long and rigid fiber threads that fail to surround cell structures (Fig. (I F) as $\nu_d$ increases. (II) Cell cluster mean elongation $E$ (Fig. (II A)), mean cell cluster number $N_C$ (Fig. (II B)), fiber cluster mean alignment $A$ (Fig. (II C)) and fiber cluster mean elongation $\Lambda$ (Fig.  (II D)) as functions of the unlinking frequency $\nu_d$ for two different values $c_1 = 0.01$ (black curve) and $c_1 = 10$ (red curve) of the flexural modulus.  \label{c1}}
  \end{figure}

Here, we perform a statistical analysis of the influence of the fiber flexural modulus $c_1$. Fig. \ref{c1} (I) shows simulations with random insemination and two different flexural moduli $c_1 = 0.01$ (first row) and $c_1 = 10$ (second row), for $\chi_\ell = 0.35$ and three different unlinking frequencies $\nu_d = 10^{-3}$ (Fig. \ref{c1}  (I A) and (I D)), $\nu_d=10^{-2}$ (Fig. \ref{c1}  (I B) and (I E)) and $\nu_d=0.2$ (Fig. \ref{c1}  (I C) and (I F)). Fig. \ref{c1} (II) shows the values of the mean elongation of cell clusters $E$ (Fig. \ref{c1}  (II A)), cell cluster number $N_C$ (Fig. \ref{c1}  (II B)), mean alignment of fiber clusters $A$ (Fig. \ref{c1}  (II C)) and mean elongation of fiber clusters $\Lambda$ (Fig. \ref{c1}  (II D)), averaged over 10 simulations and plotted as functions of the unlinking frequency $\nu_d$ for two different values $c_1 = 0.01$ (black curve) and $c_1 = 10$ (red curve) of the flexural modulus. 
 
The first row of Figs. \ref{c1} (I)  (Fig. \ref{c1}  (I A), (I B) and (I C)) shows that for a small flexural modulus $c_1=0.01$, the cell structures change from well-separated lobule-like cell clusters (Fig. \ref{c1}  (I A)) to slightly more elongated clusters (Fig. \ref{c1}  (I C)) and the fiber structures change from a disorganized fiber network (Fig. \ref{c1}  (I A) or (I B)) to more aligned fiber clusters (Fig. \ref{c1}  (I C)), as $\nu_d$ increases. For a large flexural modulus $c_1=10$ (second row of Fig. \ref{c1} I, i.e. Fig. \ref{c1}  (I D), (I E) and (I F)), the fiber structures are more aligned: the model generates organized fiber clusters able to bend around the cell structures (Fig. \ref{c1}  (I D) and (I E)). For a fast linking-unlinking (Fig. \ref{c1}  (I F)),  a rigid fiber network composed of long fiber threads which fail to surround the cell clusters is observed. 

These observations are confirmed by the values of the Statistical Quantifiers $E$ and $N_C$ (for cell clusters) and $A$ and $\Lambda$ (for fiber clusters) shown in Figs. \ref{c1} (II) as functions of the unlinking frequency $\nu_d$, for $c_1=0.01$ (black curves) and $c_1 = 10$ (red curves). Fig. \ref{c1} (II A) shows that the flexural modulus $c_1$ does not seem to significantly change the mean cell cluster elongation. Fig. \ref{c1} (II B) reveals that the number of cell clusters $N_C$ is significantly lower for $c_1=10$ than for $c_1=0.01$ for $\nu_d > 10^{-3}$. For $c_1=10$ and $\nu_d \in [10^{-3}, 0.1]$, this is because ECM rigidity is too large and the fibers fail to separate cell structures, compared to the case $c_1=1$ (see section \ref{Main:Results} of the main text). For $\nu_d>0.1$, we recover the previously described case of a fast fiber linking-unlinking dynamics. As fibers fastly self-organize into long and directed rigid threads, they force the cells to group into chord-like unseparated structures.

Figs. \ref{c1} (II C) shows that the mean alignment of the fiber clusters increases with $c_1$ (compare the black and red curves), and the difference between the values of $A$ for $c_1=0.01$ and for $c_1=10$ increases with the fiber unlinking frequency $\nu_d$. This is because the fibers are more rigidly maintained with a slow linking-unlinking dynamics than with a fast one (see previous section), and are thus less sensitive to alignment. Fig. \ref{c1} (II D) shows that the mean fiber cluster elongation $\Lambda$ is a non monotonous function of $\nu_d$ for $c_1=0.01$  (black curve)  and a monotonically decreasing function for $c_1=10$ (red curve).  For $c_1=0.01$, $\Lambda$ first increases to reach a maximal value at $\nu_d \approx 0.05$ and then decreases.  This reflects the ability that the fibers have to surround the cell clusters when the flexural modulus is small and the unlinking frequency $\nu_d$ is moderate, as already seen in the previous section. This ability is lost with a larger flexural modulus and $\Lambda$ becomes just a decreasing function of $\nu_d$. 

To sum up, large flexural modulus favors fiber alignment and ECM rigidity compared to small $c_1$. Moreover, the choice of the flexural modulus has to be carrefully linked to the fiber linking-unlinking dynamics, which also triggers fiber network alignment and rigidity.  For a well calibrated fiber linking-unlinking process and increasing values of $c_1$, the structures change from (a) compact middle sized cell clusters in a disorganized fiber network ($c_1=0.01$), (b) compact middle sized cell clusters in an organized fiber network ($c_1=1$) and (c) elongated cell clusters in fewer quantities inside an organized network ($c_1=10$). For $c_1 < 1$, ECM alignment is small and the fiber network cannot easily organize. By contrast, when $c_1=1$, ECM alignment is moderate and the fibers that are not too constrained can align. However for $c_1>1$, ECM rigidity is too large and this results in elongated cell clusters. Experimentally, it is observed that the lobules are more elongated at the periphery of the tissue than inside. Thus, our results suggest that fibers could be more stretched at the periphery. To support this hypothesis, it would be interesting to develop an experimental quantification method to estimate a local stress tensor similar to what has previously been done for adipocyte stiffness \cite{Shoham_2014}. This analysis suggests that the different morphologies observed in adipose tissues of healthy mice according to the location of fat (central or peripheral)can be due to ECM stiffness. Varying its value is sufficient to break the architecture of the tissue.

\subsection{Influence of biased insemination}\label{App:EB}
{ Figs. \ref{supEB} (I) shows simulation results obtained with random insemination (Figs. \ref{supEB} (I A), (I B)) and with biased insemination for $\alpha=10^{-1}$ (Figs. \ref{supEB} (I C), (I D)), for two values of the unlinking frequency $\nu_d$: $\nu_d = 10^{-4}$ (Figs. \ref{supEB} (I A), (I C)) and $\nu_d = 0.1$ (Figs. \ref{supEB} (I B), (I D)). We refer the reader to Fig. 3 of Main text for simulations with random and biased insemination using intermediate values of $\nu_d$ (namely $\nu_d = 10^{-3}$ and $\nu_d = 2 10^{-2}$). Fig. \ref{supEB} (II) shows the values of the mean elongation of cell clusters $E$ (Fig. \ref{supEB}  (II A)), the number of cell clusters $N_C$ (Fig. \ref{supEB}  (II B)), the mean alignment of fiber clusters $A$ (Fig. \ref{supEB} (II C)) and the mean elongation of fiber clusters $\Lambda$ (Fig. \ref{supEB}  (II D)), averaged over 10 simulations, as functions of the unlinking frequency $\nu_d$, for random insemination (in black) and for biased insemination with $\alpha = 10^{-1}$ (in red). In these simulations, the flexural modulus between linked fibers is $c_1=1$ and the linked fiber fraction $\chi_\ell = 0.35$. The numerical and model parameters can be found in Table \ref{table3}. 
\begin{figure}[H]
\centering
\includegraphics[scale=0.23]{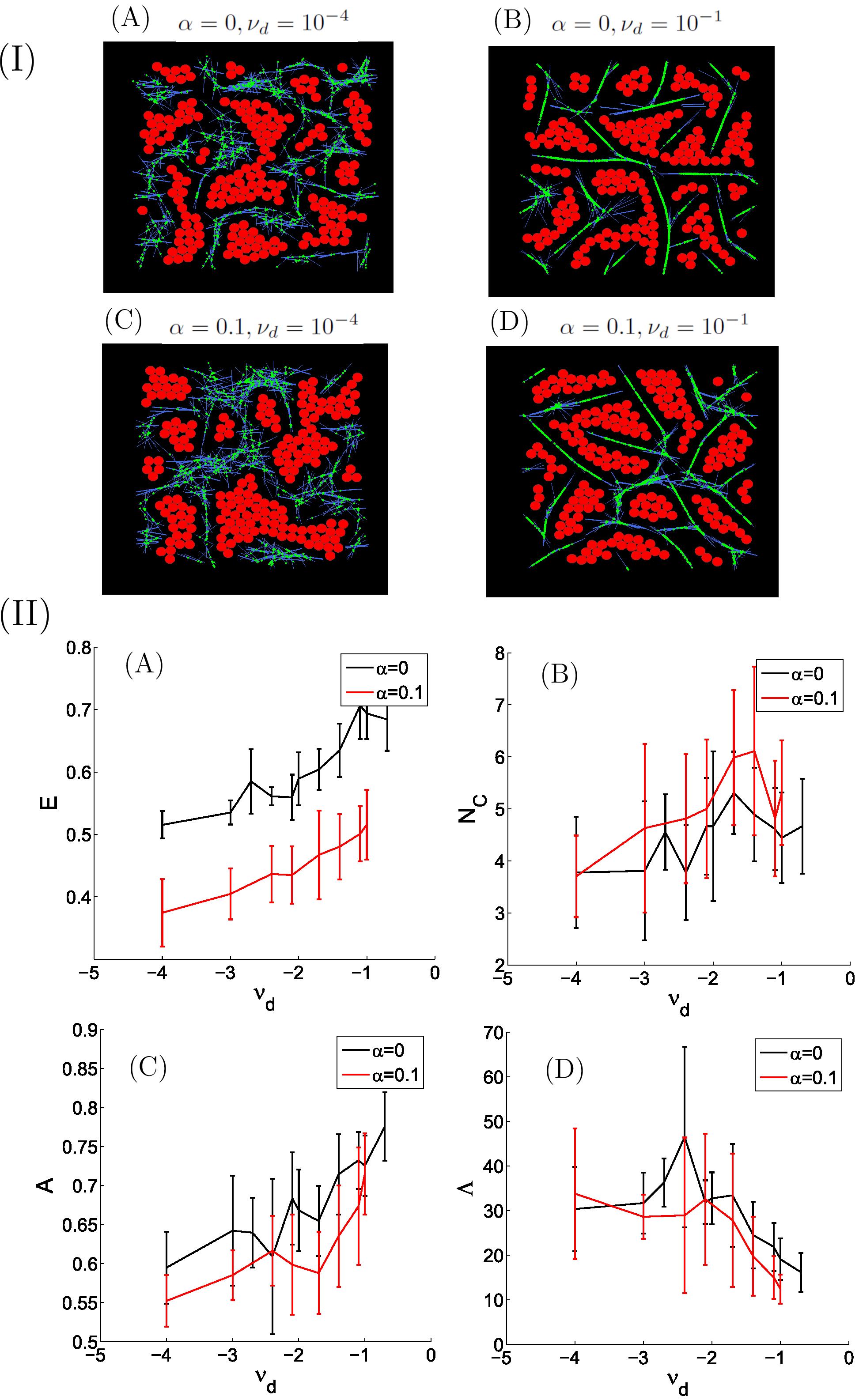}
\caption{{\bf Influence of biased insemination.} (I) Simulations with random insemination (Figs. (I A), (I B)) and with biased insemination with $\alpha=10^{-1}$ (Figs.  (I C), (I D)), for two values of the unlinking frequency $\nu_d$: $\nu_d = 10^{-4}$ (Figs. (I A) and (I C)) and $\nu_d = 0.1$ (Figs. (I B) and (I D)). Values of the flexural modulus $c_1 = 1$ and linked fiber fraction $\chi_\ell = 0.35$ have been used. (II) Cell cluster mean elongation $E$ (Fig. (II A)), mean cell cluster number $N_C$ (Fig. (II B)), fiber cluster mean alignment $A$ (Fig. (II C)) and fiber cluster mean elongation $\Lambda$ (Fig.  (II D)) averaged over 10 simulations as functions of the unlinking frequency $\nu_d$ with random insemination (in black) or with biased insemination with $\alpha=10^{-1}$ (in red). }
  \label{supEB}
  \end{figure}

Figs. \ref{supEB} (I C) and (I D) show that biased insemination leads to the creation of bigger cell clusters than random insemination (compare with Figs. \ref{supEB} (I A) and (I B)). This is because biased insemination favors insemination of new adipocytes at locations where cell clusters pre-exist. The regroupment of cells into clusters leaves regions devoid of cells (see Figs. \ref{supEB} (I C)). For cell clusters, Fig. \ref{supEB}  (II A) shows that biased insemination seems to reduce cell cluster elongation. This can be explained by the fact that cell clusters are larger with biased insemination than with random insemination, which results in a decrease of $E$. In this case, the statistical quantifier $E$ does not allow us to conclude on the form of the cell clusters. Fig. \ref{supEB} (II B) shows that biased insemination do not significantly change the number of cell clusters. Finally, Figs. \ref{supEB}  (II C) and (II D) show that biased insemination does not have a significant influence on the fiber cluster alignment $A$, but seems to slightly reduce the fiber elongation. 

Altogether, this analysis demonstrates that biased insemination with $\alpha=0.1$ does not have a significant impact on the cell and fiber structures for a properly chosen fiber linking/unlinking dynamics. This suggests that, in a sufficiently rigid fiber network, cell clusterization is mainly driven by cell-fiber interactions. Cells and fibers self-organize into middle-sized well-separated cell clusters and aligned fiber structures whatever the type of insemination (random or biased with small $\alpha$) is.}

\subsection{Anisotropic initial fiber network}\label{App:Aniso}

\begin{figure}
\centering
\includegraphics[scale=0.35]{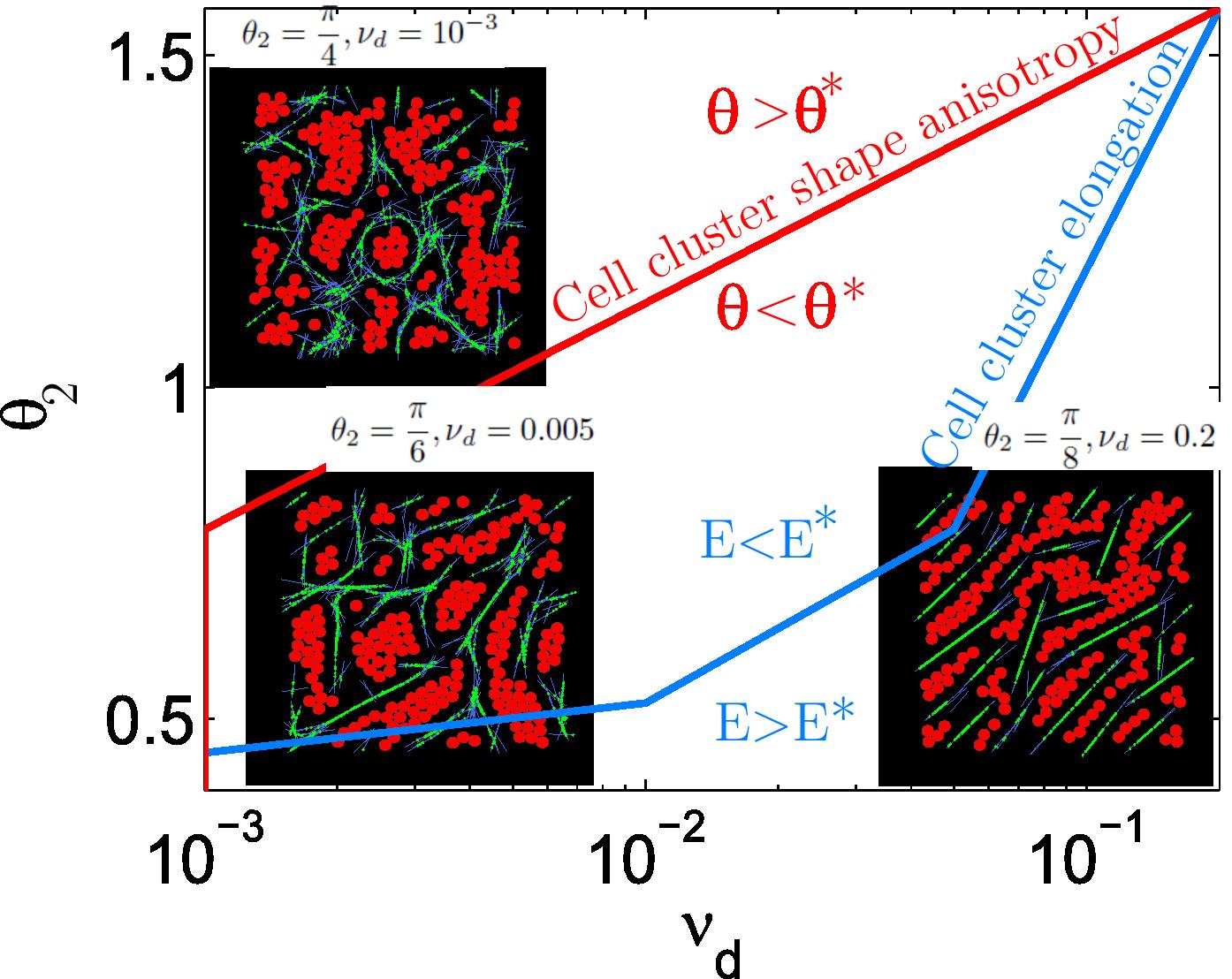}
\caption{{\bf Influence of an anisotropic initial fiber network.} According to the values of the model parameters $\nu_d$ and $\theta_2$, three phases have been obtained and classified by means of the mean cell cluster elongation $E$ and the standard deviation of cell cluster shape anisotropy $\Theta$: Phase (a): for $\Theta>\Theta^*$ and $E<E^*$. Phase (b) for $\Theta<\Theta^*$ and $E<E^*$. Phase (c) for $\Theta<\Theta^*$ and $E>E^*$. This figure displays the phase diagram in the $(E,\Theta)$ plane. Each point in this phase diagram correspond to statistical quantifiers $(E,\Theta)$ averaged over 10 simulations. The red and blue lines correspond to the separatrix between phases (a) and (b) (of equation $\Theta = \Theta^*$) and between phases (b) and (c) (of equation $E=E^*$) respectively. The figure also displays a typical simulation result for each phase, and its position on the phase diagram according to the values of $E$ and~$\Theta$. The simulations were performed  with random insemination, linked fiber fraction $\chi_\ell = 0.35$, fiber flexural modulus $c_1=1$.}
\label{ANISO}
\end{figure}

As discussed in section \ref{Main:Results} of the main text, parts of the adipose tissue reveal an anisotropic cell and fiber organization. In order to obtain oriented cell clusters, we studied the properties of the model starting from an initially anisotropic fiber network. For this purpose, we let the initial fiber directional angles $\theta^0_{f}$ be randomly chosen according to a uniform distribution in the interval $[\theta_1-\theta_2, \theta_1 + \theta_2]$, where $\theta_2$ is related to the standard deviation of the distribution. Note that the smaller $\theta_2$, the more aligned the fibers initially are. By contrast, the simulations shown so far correspond to a fully isotropic initial network, i.e. to the case $\theta_2=\pi/2$. The initial number of fiber links was carefully adjusted to be independent of the initial value of $\theta_2$ throughout the forthcoming simulations. Indeed, the probability that pairs of fibers intersect is much smaller in an aligned network than in a fully isotropic one. Simulations of Fig. \ref{ANISO} have been obtained with random insemination, flexural modulus $c_1=1$ and linked fiber fraction $\chi_\ell = 0.35$, for different unlinking frequencies $\nu_d$ and different values of $\theta_2$. Three types of structures have obtained according to the values of $\nu_d$ and $\theta_2$: (a) lobule-like non oriented cell clusters, (b) lobule-like oriented cell clusters, and (c) elongated and oriented cell clusters.  In order to quantify the passage from one morphology to another one, we use the following SQ: the mean cell cluster elongation $E$ and the standard deviation of cell cluster shape anisotropy $\Theta$. We identify the threshold values $E^*=0.65$ for $E$ (the same value as in section \ref{Main:Results} of the main text) and $\Theta^* = 0.7$ for $\Theta$. Structures of type (a) correspond to $\Theta>\Theta^*$ and $E<E^*$. Type (b) is described by $\Theta<\Theta^*$ and $E<E^*$ and finally type (c) by $\Theta<\Theta^*$ and $E>E^*$. Fig. \ref{ANISO} shows a phase diagram in the $(E,\Theta)$ plane. Each point in this phase diagram corresponds to statistical quantifiers $(E,\Theta)$ averaged over 10 simulations. The red and blue lines correspond to the separatrix between phases (a) and (b) (of equation $\Theta = \Theta^*$) and between phases (b) and (c) (of equation $E=E^*$) respectively. Fig. \ref{ANISO} also shows a typical simulation result for each phase, and its position on the phase diagram according to the values of $E$ and~$\Theta$.

Fig. \ref{ANISO} shows that, when the initial fiber network is anisotropic, the emergence of a shape anisotropy of the cell clusters depends on the fiber linking-unlinking dynamics. For a slow linking-unlinking dynamics ($\nu_d = 10^{-3}$) the initial orientation of the fibers must be strongly biased to obtain directionality in the cell and fiber final structures. Otherwise (for $\theta_2>\frac{\pi}{5}$), the initial orientation of the network is lost, and cell structures without preferential direction are obtained (see Fig. \ref{ANISO}  A). This suggests that for this slow linking-unlinking frequency, cells disturb the initial organization of the fiber network so much that this initial organization is lost. For a fast fiber linking-unlinking dynamics ($\nu_d=0.2$ see Fig. \ref{ANISO} C), cell structures are elongated due to the rigidity of the fiber network induced by the fast linking frequency and the action of the alignment torque at the created links. In this case, the fiber network imposes its preferred direction to cell cluster growth and we recover elongated cell clusters as in the case of an initially isotropic fiber network (see section \ref{Main:Results} of the main text). Fig. \ref{ANISO} (B) shows that there exist a range of values of $\nu_d$ and $\theta_2$ for which the model is able to generate lobule-like cell clusters having anisotropic shapes and a preferred shape anisotropy direction.  These configurations are obtained for $\nu_d \in ]10^{-3}, 10^{-2}[$ and for $\theta_b \in [\frac{\pi}{8},\frac{\pi}{4}]$.
 
{
  \section{Shorter fibers, smaller cell-fiber interaction range, fiber-fiber repulsion potential}\label{App:smallFib}
  
  In this section, we consider the combined effects of shortening the fiber element length, decreasing the cell-fiber interaction range and introducing a fiber-fiber repulsion potential. As described at the end of Section \ref{Main:Model} of the main text, this leads to a more biologically relevant range of parameters and phenomena. On the other hand, computing with short fiber elements necessitates a large number of such elements and is extremely time consuming. Due to these computational constraints, the simulation of Main Text were run with longer fibers, larger cell-fiber interaction range and no fiber-fiber repulsion potential.	The goal of this section is to support the relevance of this approach by showing that the results obtained in the present section with a more biologically relevant set of parameters and phenomena are similar to those described in the Main Text. That a large cell-fiber interaction range may have a similar effect as a smaller range combined with fiber-fiber repulsion makes reasonable sense. Fibers group in clusters. If they undergo a repulsion potential, these clusters are wider and affect the cells in a similar way as if the clusters were thinner (which is the case without fiber-fiber repulsion), but the cell-fiber interaction distance is larger, leading to about the same separation distance between cell clusters. 
	
	When using shorter fiber elements, we must be careful that initialization matters if we wish to compare them with longer fiber elements. Also, when the fiber element is shortened, its width encoded in the distance $d_0$ (for cell-fiber interaction) must be decreased in the same proportion. Therefore, dividing the fiber length by two indeed means replacing a fiber element by four elements connected altogether to form a diamond-shaped structure like in Fig. \ref{SIdiamond}. So, when using short fibers, we will randomly initialize diamonds like in Fig. \ref{SIdiamond}. However, as soon as the simulation is started the fiber element links in the diamonds may be removed as a consequence of the unlinking process. So, the diamond shape structures may be dissolved or persist according to how fast the unlinking process is. Considering an initial fiber network consisting of fibers longer than cell diameters is biologically plausible as collagen fibers may be produced before any implantation of adipocytes. Indeed, in wound healing, it is well documented that the wound is first populated by fibers before the appearance of the first new cell. Fiber elements in our model are not intended to reproduce a whole biological fiber but they rather produce a discretization of such fibers into rigid elements that can be computationally manipulated. At initialization actual biological fibers may be too long to be represented by a single short fiber element and need to be described by structures made of pre-assembled fiber elements. The chosen diamond shape structure is the simplest of such pre-assembled structure and it allows us to compare the results to those using large fibers as it produces similar potential isolines to the latter. 
  
  Now, to take into account fiber-fiber repulsion, we proceed as for fiber-cell repulsion as described in section \ref{App:Meca} and we suppose that the fiber repulsion potential $W_{rep}$ consists of the sum of two-particle potential elements \\ $\mathbf{W}_{rep}(Y_k,\theta_k,Y_f,\theta_f)$ modeling the mechanical interaction between fiber $f$ and fiber $k$:
  \begin{align*}
  W_{rep} = \sum_{1\leq f \leq N_f}\sum_{1\leq k \leq N_f} \big(&\mathbf{W}_{rep}(Y_k,\theta_k,Y_f,\theta_f)\\
  & + \mathbf{W}_{rep}(Y_f,\theta_f,Y_k,\theta_k) \big)
  \end{align*} For two given vectors $Y_1$, $Y_2 \in $ $\mathbb{R}^2$ and two angles $\theta_1,\theta_2 \in [-\frac{\pi}{2},\frac{\pi}{2}]$, $\mathbf{W}_{rep} = \mathbf{W}_{rep}(Y_1,\theta_1,Y_2,\theta_2)$ reads:  
\begin{equation}\label{Wrep}
  \mathbf{W}_{rep} =\begin{cases}  \frac{W_f}{d_\xi} \big(d_\xi - d(Y_1,Y_2,\theta_2)\big) \; \text{ if $d(Y_1,Y_2,\theta_2)\leq d_\xi$}\\
  0 \hspace{3,2cm}\text{ otherwise}
  \end{cases}
\end{equation}
  \noindent where $d(Y_1,Y_2,\theta_2)$ is the distance function defined by Eq. \eqref{d} and $W_f$ is a model parameter. The fiber-fiber repulsion potential element $W_{rep}$ iso-lines are ellipses with locii located at the two ends of the fiber $(Y_2,\theta_2)$. The potential vanishes beyond distance $d_\xi$ to the fiber center $Y_2$. Parameter $d_\xi$ is set such that the length of the semi minor axis is $d_0$. In this case, a fiber repels an other fiber up to distance $d_0$ in its orthogonal direction. A direct computation gives:
{
 \begin{equation*}
d_{\xi} = - L_f + 2\sqrt{\Delta_\xi}\; ,\text{  }\; \; \;  \Delta_{\xi} = (\frac{L_f}{2})^2+d_0^2.
 \end{equation*} 

We consider $N_f = 800$ fibers of length $L_f = 1.7$ (note that previously $L_f = 3$). The fiber-cell interaction range is set to $\tau R_i = R_i + d_0$ with $d_0 = 0.17$ (see Eq. \eqref{d0} and Fig. \ref{potim}). Note that previously, $d_0 = 1.32$. When fiber-fiber repulsion is activated, fibers repel their centers to a distance $d_0 = 0.17$ in their orthogonal direction, and we use the value $W_f = 10$. If not otherwise stated, the other parameters correspond to the simulations of the main text. We consider an initial fiber network composed of $N$ groups of $4$ fibers organized in 'diamond' configurations (see Fig. \ref{SIdiamond}). The center of gravity of the diamonds is chosen randomly with uniform distribution in the computational domain, and the major axis orientation of the diamonds is randomly chosen in $[-\pi, \pi]$ with uniform probability.    

\begin{figure}[h!]
\centering
\includegraphics[scale=0.2]{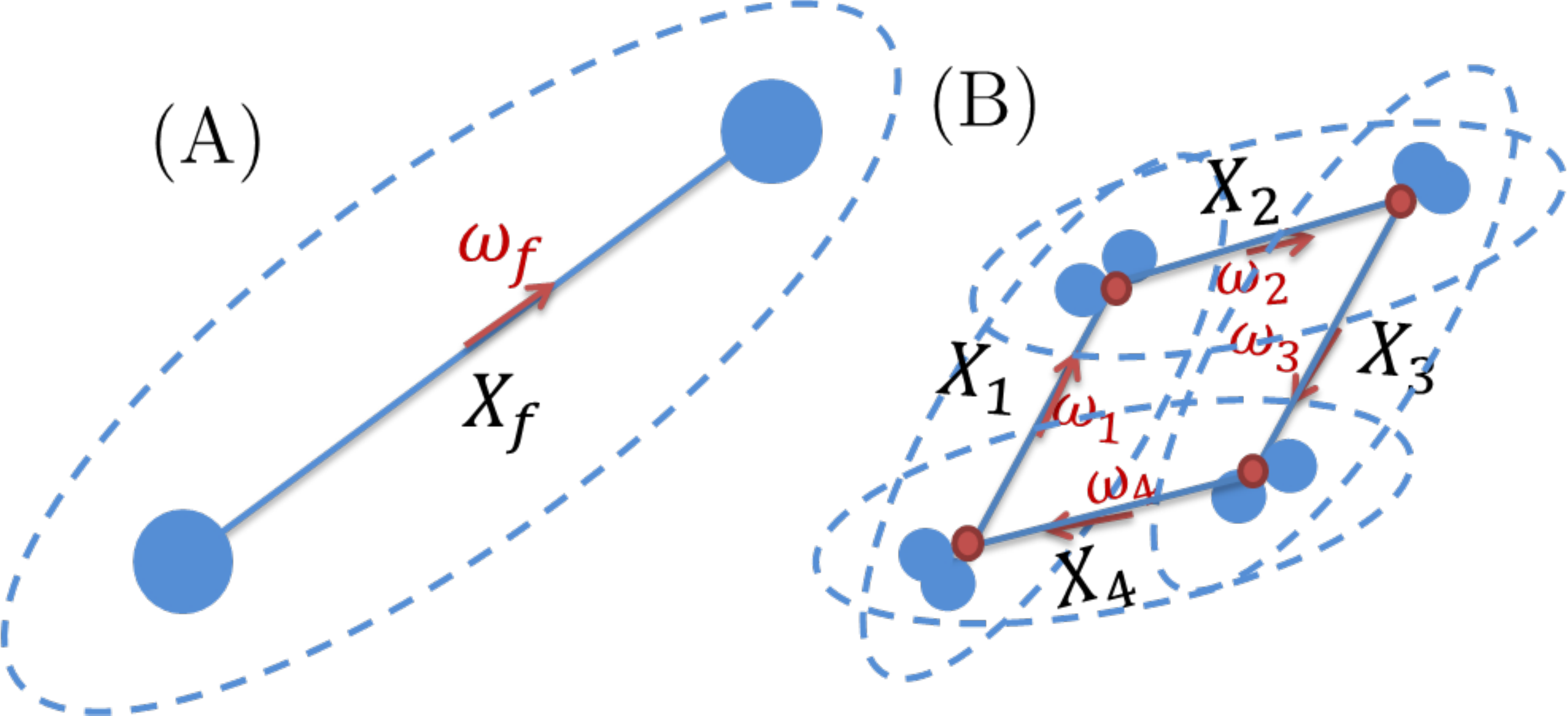}
\caption{(A) A long fiber with orientation $\omega_f$ and center $X_f$. Its repulsion zone is indicated by a dashed line. (B) Diamond configuration of 4 fibers $(X_1,\omega_1,X_2,\omega_2,X_3,\omega_3,X_4, \omega_4)$ corresponding to (A), with repulsion zones indicated by dashed lines. The two ends of the fibers are indicated by 2D-spheres and the red dots represent initial fiber links.}
\label{SIdiamond}
\end{figure}

In Fig. \ref{SIFFCI} (I), we show simulation results for $N_f = 800$, linked fiber fraction $\chi_\ell = 0.35$, small fiber unlinking frequency $\nu_d = 10^{-3}$ for two different fiber lengths: $L_f = 1.7$ (Figs. \ref{SIFFCI} (I A) and (I B)), $L_f = 1.2$ (Figs. \ref{SIFFCI} (I C) and (I D)) with random initial fiber network (Figs. \ref{SIFFCI} (I A) and (I C)) and with initial diamond configuration (Figs. \ref{SIFFCI} (I B) and (I D)). In Figs. \ref{SIFFCI} (II), we show the values of the statistical quantifiers $E$ (cell cluster elongation (II A)), $A$ (fiber alignment (II B)), $\Lambda$ (fiber cluster elongation (II C)) and $N_C$ (number of cell clusters (II D)), as functions of $\nu_d$ for $L_f=1.7$ and $L_f = 1.2$ (black and red curves), with random initial fiber network (continuous lines) and with diamond initial configuration (dashed lines).
\begin{figure}[h!]
\includegraphics[scale=0.3]{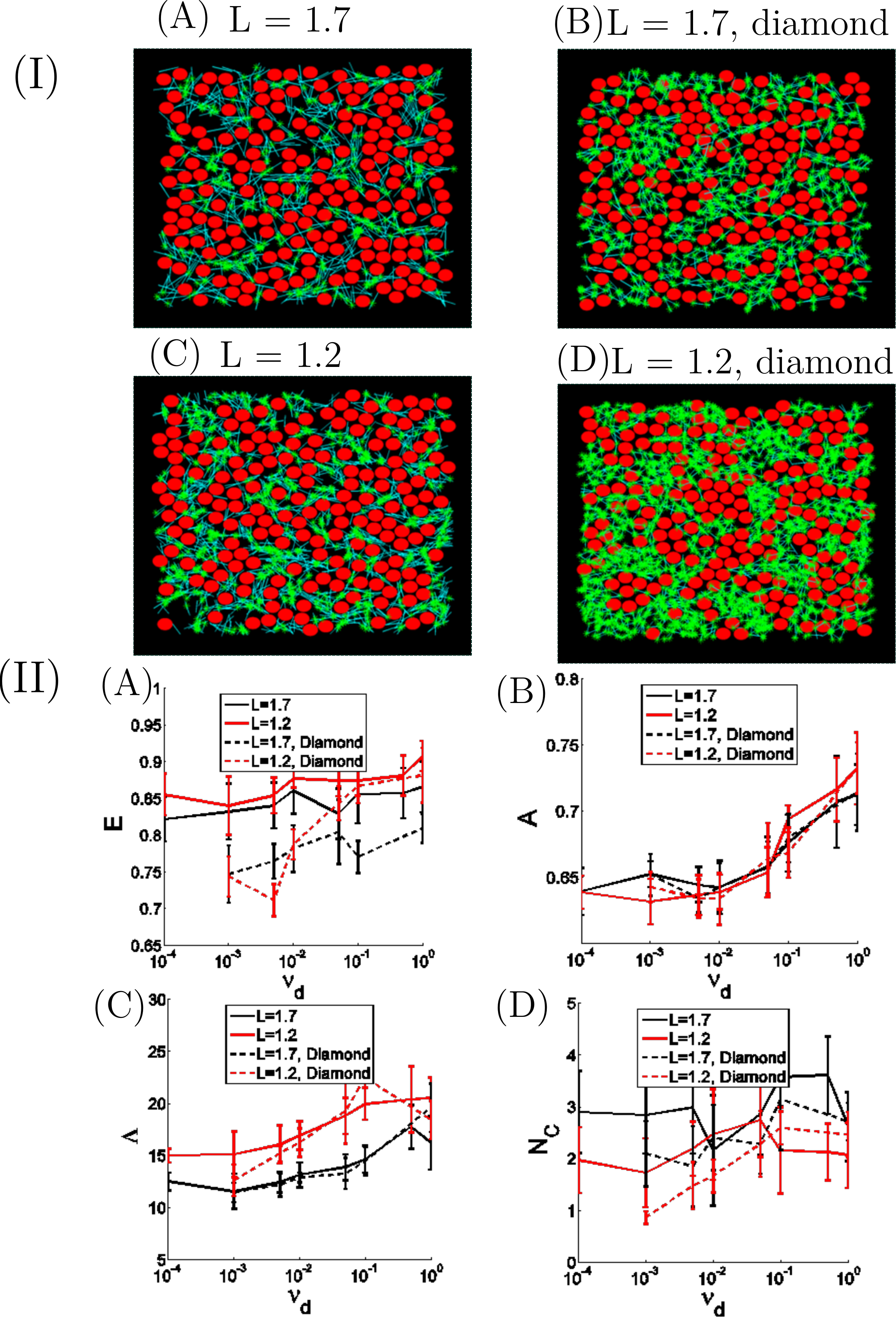}
\caption{(I) Simulation results for $N_f = 800$, linked fiber fraction $\chi_\ell = 0.35$, fiber unlinking frequency $\nu_d = 10^{-3}$ for two different fiber lengths: $L_f = 1.7$ ((I A) and (I B)), $L_f = 1.2$ ((I C) and (I D)) with random initial fiber network ((I A) and (I C)) and with initial diamond configuration ((I B) and (I D)). (II) Values of the statistical quantifiers $E$ (cell cluster elongation (II A)), $A$ (fiber alignment (II B)), $\Lambda$ (fiber cluster elongation (II C)) and $N_C$ (number of cell clusters (II D)), as function of $\nu_d$ for $L_f=1.7$ and $L_f = 1.2$ (black and red curves), with random initial fiber network (continuous lines) and with diamond initial configuration (dashed lines). \label{SIFFCI}}
\end{figure}

As shown by Figs. \ref{SIFFCI} (II), the type of initial fiber configuration does not seem to influence the fiber network (compare continuous and dashed lines of Figs. \ref{SIFFCI} (II B and C)) but has a strong influence on the cell structures. Indeed, for $L_f = 1.2$, the mean elongation of cell clusters $E$ undergoes a phase transition as $\nu_d$ increases when the initial fiber network is composed of diamond structures. This phase transition is not observed when the fibers are initially randomly distributed. This is due to the fact that for small $\nu_d$, due to its internal structure, the diamond like network is less compliant than the random one. As a result, the former exerts more pressure on the cells than the latter, favoring the emergence of small and round clusters of cells. For large $\nu_d$, i.e fast remodelling of the fiber network, the internal diamond structure is quickly lost, and we recover the same values of the SQ for diamond and random initial networks.  

To sum up, the internal diamond structure of the fiber network increases its rigidity compared to a random initial configuration. Moreover, there exists a critical fiber unlinking frequency for which the fiber network imposes local directional constraints to cell cluster growth, favoring cell cluster elongation. With such an initial diamond structure and $L_f = 1.2$, we recover the results of the main text with a random initial network and $L_f = 3$. Note however that the fiber network is less organized with small fibers than with larger ones (compare the values of $A$ and $\Lambda$ between Figs. \ref{SIFFCI}, Fig. 2 of the main text and \ref{mainSIMUS}). This is due to the fact that the fibers of the diamonds are not strictly aligned, and this effect adds up with the fiber-fiber repulsion potential which disorganizes the fiber network. 

Finally, we show the influence of the linked fiber fraction when fiber-fiber repulsion is activated, for $L_f = 1.2$ and with diamond-like initial configuration for the fiber network. In Fig. \ref{SIFFchi}, we show contour plots of the SQ $E$ (cell cluster elongation, Plot A), $A$ (fiber alignment, Plot B), $\Lambda$ (fiber cluster elongation, Plot C) and $N_C$ (number of cell clusters, Plot D), obtained when varying the linked fiber fraction $\chi_\ell$ (vertical axis) and the fiber unlinking frequency $\nu_d$ (horizontal axis).
 
\begin{figure}[h!]
\includegraphics[scale=0.5]{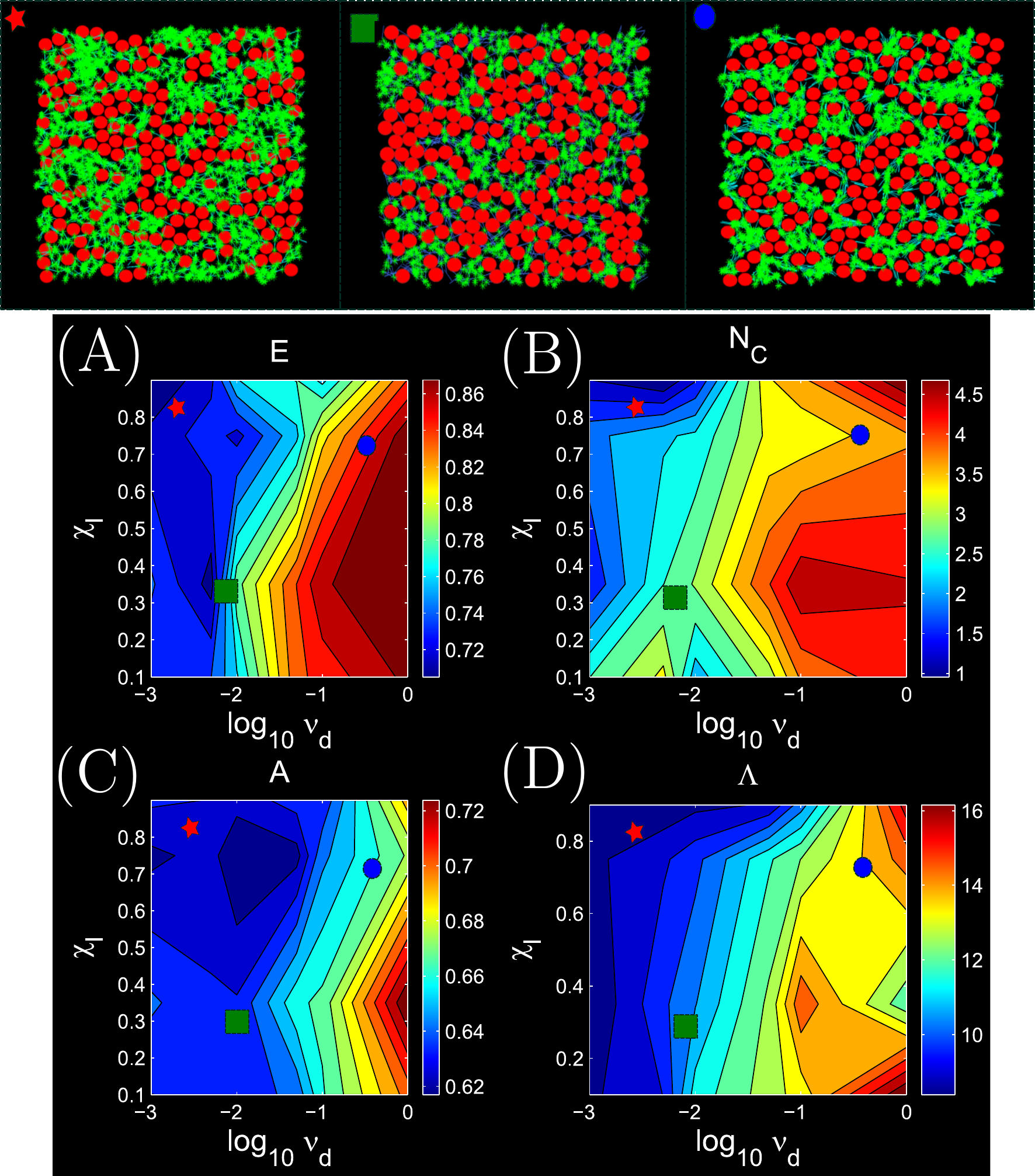}
\caption{Contour plots of the statistical quantifiers $E$ (cell cluster elongation, A), $N_C$ (number of cell clusters, B), $A$ (fiber alignment, C), and $\Lambda$ (fiber cluster elongation, D), obtained when varying the fiber linked fraction $\chi_\ell$ (vertical axis) and the fiber unlinking frequency $\nu_d$ (horizontal axis). The parameter values $\nu_d$ and $\chi$ of Figures indicated by {\large ${\color{red} \bigstar}$}, {\large ${\color{green} \blacksquare}$} and {\large ${ \bullet}$} are shown on the graph and correspond to $(\nu_d, \chi) = (10^{-3}, 0.9)$, $(\nu_d, \chi) = (10^{-2}, 0.35)$ and $(\nu_d, \chi) = (0.2, 0.75)$ respectively. \label{SIFFchi}}
\end{figure}

As shown in Fig. \ref{SIFFchi}, we recover the morphologies observed in the main text (A) lobule-like cell clusters surrounded by a disorganized (unaligned) fiber network, (B) lobule-like cell structures in an aligned fiber network, and (C) elongated cell structures in a network composed of
long and rigid fiber threads. We also recover the phase transitions in $\nu_d$ and $\chi_\ell$ for the statistical quantifiers $E$ and $A$ that we observed in Fig. 2 of the main text (with no fiber-fiber repulsion, $L_f = 3$ and random initial network). The most biologically relevant structures, composed of well-separated lobule-like cell clusters in an organized fiber network, are here again identified around $\nu_d \in [10^{-2}, 10^{-3}]$ for $\chi_\ell \geq 0.35$. Note however that the fiber network is less aligned than in the simulations of the main text.

To sum up, adding a fiber-fiber repulsion potential does not lead to a breakdown of the results previously obtained without such a repulsion potential. On the contrary, it reinforces them by making them more robust to parameter changes. These results show that the fiber network of the main text can be seen as a fiber network composed of initially connected smaller fibers which repulse each other. 
}

}
 
 {
 \section{Computational time of the model}
 
The individual based model was coded in FORTRAN90 in sequential mode for each simulation, and sets of simulations were run in parallel for the statistical analysis. The computation time was optimized by the use of a numerical grid to reduce the size of the computational domain around each agent, and one simulation took from few hours to several days as functions of the model parameters. In Fig. \ref{App:CPU}, we show the computational times -in hours- for simulations of Appendix F, i.e when considering fiber-fiber repulsion, for $\chi_\ell = 0.1$ and three different values of $\nu_d$: $\nu_d = 10^{-3}$ (blue curves), $\nu_d = 10^{-2}$ (red curves) and $\nu_d = 0.1$ (yellow curves). The computational time corresponds to the time a simulation takes to reach the steady state computed at $t=80t_{ref}$, and is measured as the difference between the times at which the data files of the final and initial states are written by the program in the data folder. We plot the computational time as a function of the fiber total area (top figure), and of the number of fiber links (bottom figure). The fiber total area $A_T$ corresponds to the area the fibers would occupy if it was in optimal (non-overlapping) configuration, i.e it is computed as the sum of all fiber areas:
$$
A_T = N_f d_0 \frac{L_f}{2},
$$
\noindent where $d_0$ is the length of the semin minor axis of the ellipse around fiber $f$, and $\frac{L_f}{2}$ the length of its semi major axis (see Appendix F). Simulations of Figs. \ref{App:CPU} are produced with varying the number of fibers $N_f = \{200,300,400$ $,500,600,800,1000\}$, the fiber length $L_f = \{1.2,2.4,3\}$ and $d_0 = \{0.12,0.24,0.3\}$.
\begin{figure}[h]
\includegraphics[scale=.2]{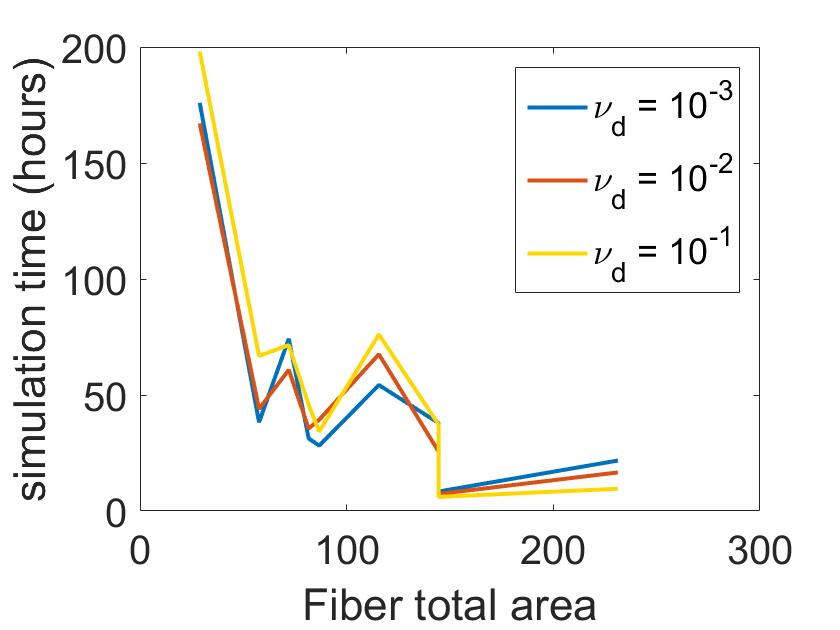}
\includegraphics[scale=.2]{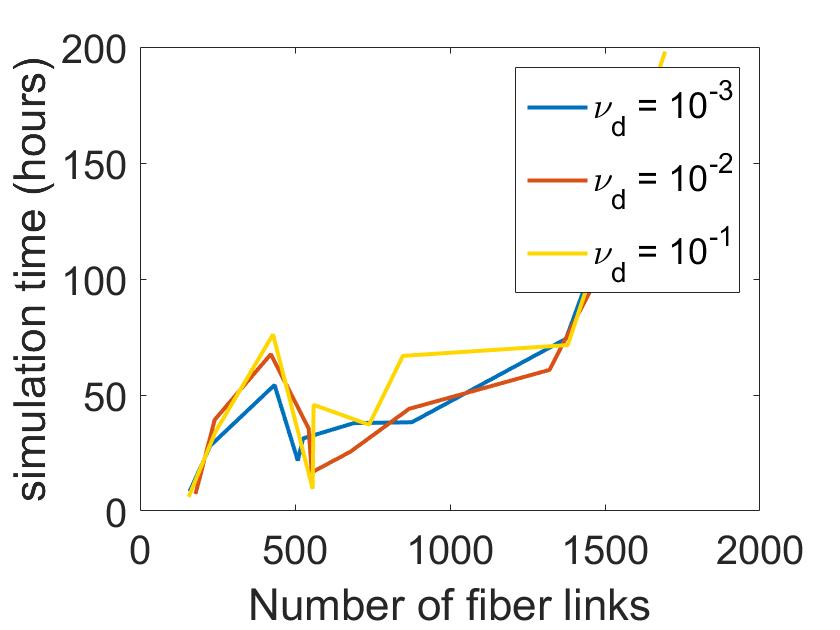}
\caption{Computational time -in hours- for simulations of Appendix F with $\chi_\ell = 0.1$ and three different values of $\nu_d$: $\nu_d = 10^{-3}$ (blue curves), $\nu_d = 10^{-2}$ (red curves) and $\nu_d = 0.1$ (yellow curves). Top: as a function of the fiber total area. Bottom: as a function of the number of fiber links.\label{App:CPU}}
\end{figure}
\noindent As shown in Figs. \ref{App:CPU}, the computational time decreases with the fiber total area $A_T$ for all values of $\nu_d$, while it increases as a function of the number of fiber links in an almost linear way. This shows that the computational time depends non trivially on the model parameters. As the total area occupied by the fibers increases (by increasing either the number of fibers either each fiber repulsion area), the tissue becomes more and more constrained by the fiber network, and displays fewer degrees of freedom: fibers and cells have less space to move. As a result, the system is closer to its optimal state at each discrete time of the simulation, and this can result in a decrease of the number of iterations in the minimization algorithm. We stress the fact that the computational time displayed in Figs. \ref{App:CPU} is the total time of a simulation. This includes the time to compute the interactions between each pair of interacting agents, as well as the time to perform the minimization process between each time steps. As a result, this CPU time is not straightforwardly linked to the number of agents as could be expected. Here, the more constrained the system, the less self-organization is possible, resulting in a small computational time. Finally in the bottom figure of Fig. \ref{App:CPU}, we show the computational time of the model as a function of the total number of fiber links in the model. As expected, the computational time increases with the number of fiber links, in an almost linear way. This reflects the larger time needed to compute the interactions when increasing the number of fiber links. The simulations of the main text or appendix F correspond to a fiber area approximately equal to $50$, with a mean computational time of 50 hours for $800$ fibers of length $L_f = 1.7$.

}
%\section{Three-dimensional organization of adipose tissue}\label{App:3D}

\section{Supplementary Informations: Movies}

\subsection{S1 Video}
\label{S1_Vid}
{\bf Simulation: Fixed number of fiber links ($\nu_d=\nu_\ell=0$), moderately constrained fiber network ($40\%$ of links on the population of crossing fibers).}  Cells (2D spheres) are represented in red, fibers (straigth lines) in blue and fiber links as green crosses. This video shows a simulation with a fixed number of fiber links, for a moderately constrained fiber network. Cells and fibers self-organize to form lobule-like structures of cells in a stiff fiber network (moderately organized). 
Biologically relevant cell and fiber structures are obtained, with a fiber network slightly more disorganized than in real tissues.

\subsection{S2 Video}
\label{S2_Vid}
{\bf Simulation: Fixed number of fiber links ($\nu_d=\nu_\ell=0$), highly constrained fiber network ($70\%$ of links on the population of crossing fibers).} Cells (2D spheres) are represented in red, fibers (straigth lines) in blue and fiber links as green crosses. This video shows a simulation with a fixed number of fiber links, for a highly constrained fiber network. Cells are repelled into zones of low fiber density and newly differentiated cells are then attracted into these zones. This results in the formation of well-separated lobule-like structures of cells. However, the fiber network is disorganized because of the large number of fixed fiber links. Biologically relevant cell structures are obtained, in a fiber network which could represent a pathological situation such as fibrosis.

\subsection{S3 Video}
\label{S3_Vid}
{\bf Simulation: Dynamical number of fiber links ($\nu_d,\nu_\ell \neq 0$), moderately constrained fiber network ($40\%$ of links on the population of crossing fibers initially), moderate linking/unlinking frequency.} Cells (2D spheres) are represented in red, fibers (straigth lines) in blue and fiber links as green crosses. This video shows a simulation with a moderate linking/unlinking frequency. Cells and fibers self-organize to form lobule-like structures of cells in an organized fiber network. The dynamical linking/unlinking favors the organization of the fiber network compared to a fixed number of fiber links (to be compared with \ref{S1_Vid}). Biologically relevant cell and fiber structures are obtained.

\subsection{S4 Video}
\label{S4_Vid}
{\bf Simulation: Dynamical number of fiber links ($\nu_d,\nu_\ell \neq 0$), moderately constrained fiber network ($40\%$ of links on the population of crossing fibers initially), fast linking/unlinking.} Cells (2D spheres) are represented in red, fibers (straigth lines) in blue and fiber links as green crosses. This video shows a simulation with fast fiber linking/unlinking. The fiber network is more rigid in this regime compared with the case of a slow linking/unlinking process. A preferred fiber direction locally emerges and generates elongated cell clusters. Such elongated structures could model other organs, such as muscles.

%\nolinenumbers


\begin{thebibliography}{10}

\bibitem{Ailhaud_1999}
Ailhaud G (1999) Cross talk between adipocytes and their precursors: relationships with adipose tissue development and blood pressure. {\it Annals of the New York Academy of Sciences} 892:127-133.

\bibitem{Alonso_2014}
Alonso R, Young J, Cheng Y (2014) A particle interaction model for the simulation of biological and cross-linked fibres inspired from flocking theory. {\it Cellular and molecular
bioengenering} 7(1):58-72.

\bibitem{Ambrosi_2005}
Ambrosi D, Bussolino F, Preziosi L (2005) A review of vasculogenesis models. {\it Journal of Theoretical Medecine} 6(1):1-19.

\bibitem{Arner_2010}
Arner E, et al. (2010) Adipocyte turnover: relevance to human adipose tissue morphology. {\it Diabetes} 59(1):105-109.

\bibitem{Chen_2001}
Chen C-Y, Byrne H-M, King J-R (2001) The influence of growth induced stress from the surrounding medium on the development of multicell spheroids. {\it J Math Biol} 43:191-220.

\bibitem{Ciarletta_2009}
Ciarletta P, Ben Amar M (2009) A finite dissipative theory of temporary interfillar bridges in the extra-cellular matrix of ligaments and tendons. {\it Interface} 6(39):909-924.

\bibitem{Corvera_2014}
Corvera S, Gealekman O (2014) Adipose tissue angiogenesis: impact on obesity and type-2 diabetes. {\it Biochim Biophys Acta} 1842(3):463-72.

\bibitem{Cowin_AnnRevBiomedEng04}
Cowin SC (2004) Tissue growth and remodeling. {\it Annu. Rev. Biomed. Eng.} 6:77-107. 

\bibitem{Crisan_etal_CellStemCell08}
Crisan M, Yap S, Casteilla L, Chen CW, Corselli M, Park TS, Andriolo G, Sun B, Zheng B, Zhang L, Norotte C, Teng PN, Traas J, Schugar R, Deasy BM, Badylak S, Buhring HJ, Giacobino JP, Lazzari L, Huard J, P\'eault B (2008) A perivascular origin for mesenchymal stem cells in multiple human organs. {\it Cell Stem Cell} 3:301-313.

\bibitem{Cristancho_2011}
Cristancho A-G, Lazar M-A (2011) Forming functional fat: a growing understanding of adipocyte differentiation. {\it Nat Rev Mol Cell Biol} 11:722-34.

\bibitem{Degond_etal_M3AS15} Degond P, Delebecque F, Peurichard D (2015) Continuum model for linked fibers with alignment interactions. {\it Math Models Methods Appl Sci}  26:269-318.  

\bibitem{Dickinson_2000}
Dickinson R-B (2000) A generalized transport model for biased cell migration in an anisotropic environment. {\it J Math Biol} 40:97-135.

\bibitem{Divoux_2011}
Divoux A, Clement K (2011) Architecture and the extracellular matrix: the still unappreciated components of the adipose tissue. {\it Obes Rev} 12:494-503.

\bibitem{Drasdo_2003}
Drasdo D (2003) On selected individual-based approaches to the dynamics in multicellular systems. {\it Multiscale Modelling and Numerical Simulations}, eds W. Alt, M.
Chaplain, M. Griebel, J. Lenz (Birkh\"auser, Basel, Switzerland), pp 109-203.

\bibitem{Drasdo_2005}
Drasdo D, H\"ohme S (2005) A single cell based-model of tumor growth in-vitro: monolayers and spheroids. {\it Phys Biol} 2(3):133-47.

\bibitem{Friedl_2000}
Friedl P, Br\"ocker E-B (2000) The biology of cell locomotion within three dimensional extracellular matrix. {\it Cell Mol Life Sci} 57(1):41-64

\bibitem{Friedl_2004}
Friedl P, Hegerfeldt Y, Tusch M (2004) Collective cell migration in morphogenesis and cancer. {\it Int J Dev Biol} 48:441-449.

\bibitem{Galanis_2010}
Galanis J , Nossala R and Harries D (2010) Depletion forces drive polymer-like self-assembly in vibrofluidized granular materials. {\it Soft Matter} 6:1026-1034.

\bibitem{Galle_2009}
Galle J, Hoffmann M, Aust G (2009) From single cells to tissue architecture : a bottom-up approach to modelling the spatio-temporal organization of complex multicellular systems. {\it J Math Biol} 58(1-2):261-83.

\bibitem{Guido_1993}
Guido S, Tranquillo R-T (1993) A methodology for the systematic and quantitative study of cell contact guidance in oriented collagen gels correlation of fibroblast orientation and gel birefringence. {\it J Cell Sci} 105:317-331.

\bibitem{Hemmingsen_2013}
Hemmingsen M, et al. (2013) The role of paracrine and autocrine signaling in the early
phase of adipogenic differentiation of adipose-derived stem cells. {\it PloS One} 8(5):e63638.

\bibitem{Hillen_2006}
Hillen T (2006) Mesoscopic and macroscopic models for mesenchymal motion. {\it J Math Biol} 53(4):585-616.

\bibitem{Hillen_2010}
Hillen T, Hinow P, Wang Z-A (2010) Mathematical analysis of a kinetic model for cell movement in network tissues. {\it Discrete and Continuous Dynamical Systems} 14(3):1055-1080.

\bibitem{Hosek_2004}
Hosek M, Tang J.X (2004) Polymer-induced bundling of F actin and the depletion force. {\it Phys Rev E} 69:051907.

\bibitem{Hwang_2009}
Hwang M, Garbey M, Berceli S-A, Tran-Son-Tay R (2009) Rule-Based Simulation of Multi-Cellular Biological Systems-A
Review of Modeling Techniques, {\it Cell Mol Bioeng} 2(3):285-294.

\bibitem{Joanny_2007}
Joanny J-F, J\"ulicher F, Kruse K, Prost J (2007) Hydrodynamic theory for multicomponent active polar gels. {\it New J Phys} 9:422.

\bibitem{Jungblut_2007}
 Jungblut S.,  Tuinier R., Binder K. and Schilling T (2007) Depletion induced isotropic-isotropic phase separation in suspensions of rod-like colloids.  {\it J Chem Phys} 127:244909.

\bibitem{Khan_etal_MolCellBiol09}
Khan T, Muise ES, Iyengar P, Wang ZV, Chandalia M, Abate N, Zhang BB, Bonaldo P, Chua S, Scherer PE (2009) Metabolic dysregulation and adipose tissue fibrosis: role of collagen VI. {\it Mol Cell Biol} 29:1575-1591. 

\bibitem{Lushnikov_2008}
Lushnikov P-M, Chen N, , Alber M (2008) Macroscopic dynamics of biological cells interacting via chemotaxis and direct contact. {\it Physic Rev E} 78:061904

\bibitem{Murray_1983}
Murray J-D, Oster G-F, Harris A-K (1983) A mechanical model for mesenchymal morphogenesis.{\it J Math Biol} 17:125-129.

\bibitem{Napolitano_1963}
Napolitano L (1963) The differentiation of white adipose cells an electron microscope
study. {\it J Cell Biol} 18:663-679.

\bibitem{Ouchi_2011}
Ouchi N, Parker J-L, Lugus J-J and Walsh K (2011), Adipokines in inflammation and metabolic disease. {\it Nat Rev Immunol} 11:85-97.

\bibitem{Pellegrinelli_etal_JPathol14}
Pellegrinelli V, Heuvingh J, du Roure O, Rouault C, Devulder A, Klein C, Lacasa M, Cl\'ement E, Lacasa D, Cl\'ement K (2014) Human adipocyte function is impacted by mechanical cues. {\it J Pathol} 233:183-95.
 
\bibitem{Peurichard_2016} Peurichard D (2016) Macroscopic model for cross-linked fibers with alignment interactions: Existence theory and numerical simulations. {\it Mult. Model. Simul.} 14-4 (2016), pp. 1175-1210.  

\bibitem{Preibisch_2009}
Preibisch S, Saalfeld S, Tomancak P (2009) Globally optimal stitching of tiled 3d microscopic image acquisitions.{\it Bioinformatics} 25(11):1463-1465.

\bibitem{Rodriguez_1994}
Rodriguez E-K, Hoger A, McCulloch A-D (1994) Stress-dependent finite growth in soft elastic tissues. {\it J Biomech} 27(4):455-467.

 \bibitem{Shraiman_2005}
Shraiman B-I (2005) Mechanical feedback as a possible regulator of tissue growth. {\it PNAS} 102(9):3318-3323.

\bibitem{Sparbati_etal_EurJHistochem10}
Sbarbati A, Accorsi D, Benati D, Marchetti L, Orsini G, Rigotti G, Panettiere P (2010) Subcutaneous adipose tissue classification. {\it Eur J Histochem} 54:e48.

\bibitem{Taber_2011}
Taber L-A, Shi Y, Yang L, Bayly P-V (2011) A poroelastic model for cell crawling including mechanical coupling between cytoskeletal contraction and actin polymerization.{\it J Mech Mat Struct} 6:569-589.

\bibitem{Tang_etal_Science08}
Tang W, Zeve D, Suh JM, Bosnakovski D, Kyba M, Hammer RE, Tallquist MD, Graff JM (2008) White fat progenitor cells reside in the adipose vasculature. {\it Science} 322:583-586. 

\bibitem{Vincent_1991}
Vincent L, Soille P (1991) Watersheds in digital spaces: An efficient algorithm based on immersion simulations. {\it IEEE Trans Pattern Anal Mach Intell} 13(16):583-598.

\bibitem{Uzawa_1958}
Arrow KJ, Hurwicz L, Uzawa H (1958) Studies in linear and nonlinear programming, Stanford University Press 

\bibitem{Shoham_2014}
Shoham N, et al. (2014) Adipocyte stiffness increases with accumulation of lipid droplets. {\it Biophys J} 106(6):1421-31.

\bibitem{Sepe_2011}
Sepe A, Tchkonia T, Thomou T, Zamboni M, Kirkland J-L (2011) Aging and regional differences in fat cell progenitors - a mini-review. {\it Gerontology} 57(1):66-75.

\bibitem{Sun_2013}
Sun K, Tordjman J, Clement K, Scherer PE (2013) Fibrosis and adipose tissue dysfunction.{\it Cell Metab} 18(4):470-477.

\bibitem{Ushiki_2002}
Ushiki T (2002), Collagen fibers, reticular fibers and elastic fibers. A comprehensive understanding from a morphological viewpoint. {\it Arch. Histol. Cytol.} 65(2):109-126 

\bibitem{Wasserman_2011}
Wasserman F (2011) The development of adipose tissue. {\it Compr physiol Supplement} 15: Handbook of Physiology, Adipose Tissue:87-100 First published in print
1965. doi:101002/cphycp050110.


\end{thebibliography}
\end{document}